\documentclass[11pt]{article}
\usepackage{booktabs} % To thicken table lines

\pdfoutput=1
\usepackage{latexsym}
\usepackage{epsfig}
\usepackage[mathscr]{eucal}
\usepackage{amsfonts}
\usepackage{amscd}
\usepackage{cite}
\usepackage{array}
\usepackage{bbold}
\usepackage{amssymb}
\usepackage{colordvi}
\usepackage[centertags]{amsmath}
\usepackage{enumerate}
\usepackage{graphicx}
\usepackage{booktabs}
\usepackage{theorem}
\usepackage{caption}
\usepackage{soul}
\usepackage{url}
\usepackage[hidelinks]{hyperref}
\usepackage{mcite}
\usepackage{slashed}
\usepackage[dvipsnames]{xcolor}
\usepackage{bbm}
\usepackage[utf8]{inputenc}
\usepackage{scrextend}
\usepackage{listings}
\usepackage{multirow}
\usepackage{textpos}

\definecolor{codegreen}{rgb}{0,0.6,0}
\definecolor{codegray}{rgb}{0.5,0.5,0.5}
\definecolor{codepurple}{rgb}{0.58,0,0.82}
\definecolor{backcolour}{rgb}{0.95,0.95,0.92}

\lstdefinestyle{mystyle}{
    backgroundcolor=\color{backcolour},   
    commentstyle=\color{codegreen},
    keywordstyle=\color{blue},
    numberstyle=\tiny\color{codegray},
    stringstyle=\color{codepurple},
    basicstyle=\ttfamily\footnotesize,
    breakatwhitespace=false,         
    breaklines=true,                 
    captionpos=b,                    
    keepspaces=true,                 
    numbers=left,                    
    numbersep=5pt,                  
    showspaces=false,                
    showstringspaces=false,
    showtabs=false,                  
    tabsize=2
}

\allowdisplaybreaks{}

\numberwithin{equation}{section}

\newcommand{\be}{\begin{equation}}
\newcommand{\ee}{\end{equation}}
\newcommand{\beq}{\begin{eqnarray}}
\newcommand{\eeq}{\end{eqnarray}}
\newcommand{\RE}{\texttt{RelExt}}

\newcommand{\RNum}[1]{\uppercase\expandafter{\romannumeral #1\relax}}

\newcommand{\si}[1]{\text}
\newcommand{\SI}[1]{\text}

\newcommand{\s}{\newline \vspace*{-3.5mm}}

\newcommand{\DT}{\texttt{RelExt}}
\newcommand{\DTpath}{\texttt{\$RelExt}}
\usepackage{comment}

\newcommand{\ttt}[1]{\texttt{#1}}

\begin{document}

\lstset{style=mystyle}

\title{

  \vspace{-1.3cm} % Falls nötig, um die Nummer weiter hoch zu setzen
    \hfill {\small \hfill KA-TP-05-2025}\\[1cm]

     \begin{minipage}{0.07\textwidth} % Linker Bereich für das Logo
        \centering
        \includegraphics[scale=0.1]{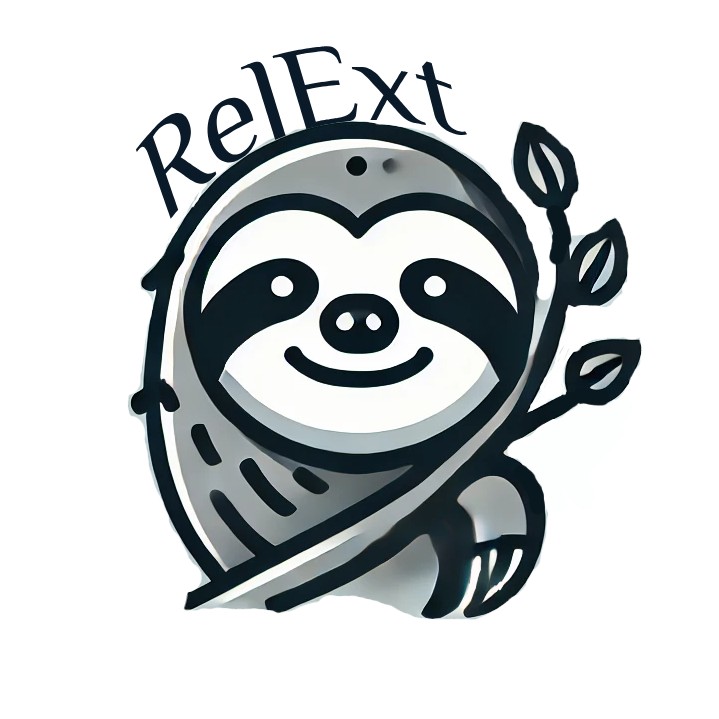} % Logo anpassen
    \end{minipage}
    \hspace{0.1cm}
    \begin{minipage}{0.75\textwidth} % Rechter Bereich für den Text
        \centering
        {\DT: A New Dark Matter Tool for the \\ Exploration of Dark Matter Models} % Untertitel
    \end{minipage}  
}

\date{\today}
\author{
Rodrigo Capucha$^{1\,}$\footnote{E-mail: \texttt{rscapucha@fc.ul.pt}},
Karim Elyaouti$^{2\,}$\footnote{E-mail:
	\texttt{karim.elyaouti@partner.kit.edu}},
Margarete M\"{u}hlleitner$^{2\,}$\footnote{E-mail:
	\texttt{margarete.muehlleitner@kit.edu}},\\
Johann Plotnikov$^{2\,}$\footnote{E-mail:  \texttt{johann.plotnikov@online.de}},
Rui Santos$^{1,3\,}$\footnote{E-mail:  \texttt{rasantos@fc.ul.pt}}
\\[9mm]
{\small\it
$^1$Centro de F\'{\i}sica Te\'{o}rica e Computacional,
    Faculdade de Ci\^{e}ncias,} \\
{\small \it    Universidade de Lisboa, Campo Grande, Edif\'{\i}cio C8
  1749-016 Lisboa, Portugal} \\[3mm]
{\small\it
$^2$Institute for Theoretical Physics, Karlsruhe Institute of Technology,} \\
{\small\it Wolfgang-Gaede-Str. 1, 76131 Karlsruhe, Germany.}\\[3mm]
{\small\it
$^3$ISEL -
 Instituto Superior de Engenharia de Lisboa,} \\
{\small \it   Instituto Polit\'ecnico de Lisboa
 1959-007 Lisboa, Portugal} \\[3mm]
}

\maketitle
\begin{abstract}
We present the {\tt C++} program $\DT$ for Standard Model (SM) extensions that feature a Dark Matter (DM) candidate. The tool allows to efficiently scan the parameter spaces of these models to find parameter combinations that lead to relic density values which are compatible with the measured value within the uncertainty specified by the user. The code computes the relic density for freeze-out (co-)annihilation processes. The user can choose between several pre-installed models or any arbitrary other model featuring a discrete $\mathbb{Z}_2$ symmetry, by solely providing the corresponding \texttt{FeynRules} model files. The code automatically generates the required (co-)annihilation amplitudes and thermally averaged cross sections, including the total widths in the $s$-channel mediators, and solves the Boltzmann equation to determine the relic density. It can easily be linked to other tools like e.g.~\texttt{ScannerS} to check for the relevant theoretical and experimental constraints, or to \texttt{BSMPT} to investigate the phase history of the model and possibly related gravitational waves signals.
\end{abstract}
\newpage

\tableofcontents

\newpage
\section{Introduction}
The true nature of Dark Matter (DM) is one of the most prominent open questions of contemporary particle physics. While the Standard Model (SM) of particle physics has been structurally completed with the discovery of the Higgs boson~\cite{Aad:2012tfa, Chatrchyan:2012ufa}, which turned out to behave very SM-like, and has been tested to the highest accuracy, it does not provide a viable DM candidate. Numerous astrophysical and cosmological observations point to the existence of DM~\cite{hoekstra2002nf,koopmans2002qh,metcalf2003sz,moustakas2002iz,rubin1970rotation,clowe2004weak}. However, it remains totally unclear what is the nature of DM.  It can range from particle character to supermassive objects like e.g.~primordial black holes~\cite{Chapline:1975ojl}, and thus cover a huge mass range. Assuming particle character for DM, there are two main mechanisms that can produce DM in agreement with all observables, in particular the measured relic density~\cite{Planck:2018vyg}. These are given by freeze-in \cite{Hall:2009bx} and freeze-out \cite{Zeldovich:1965gev,Bertone:2004pz}, where we consider here the latter process. In the freeze-out mechanism, weakly interacting massive particles (WIMPs) are assumed to be in thermal equilibrium with the thermal bath until the expansion rate of the universe becomes larger than the DM (co-)annihilation rate into lighter particles. The thermal relic density that we observe today is then obtained from the solution of the Boltzmann equation for the evolution of the DM number density. \s

A prerequisite for any model predicting DM is its compatibility with observations. Besides theoretical and other experimental constraints, the involved DM particles have to be compatible with collider search limits, constraints from direct detection and indirect detection, and last but not least they have to lead to the correct relic density. A minimum requirement is that the predicted relic density does not exceed the experimental value. In this context, models implying a relic density below the observed value may be considered as phenomenologically viable, if one assumes that there may exist extensions of the model with DM particles saturating the relic density. An interesting question, however, is whether there still exists parameter combinations of the model under investigation that lead to the measured value of the relic density within the experimental uncertainties and what their phenomenological implications would be. Depending on the complexity of the parameter space, it may turn out to be quite time consuming to find such parameter sets, and sophisticated scan procedures are required. \s 

There exist numerous codes that calculate for beyond-SM (BSM) extensions DM observables, in particular the DM relic density, such as {\tt SuperIso Relic} \cite{Arbey:2018msw}, {\tt DarkSUSY} \cite{Bringmann:2018lay}, {\tt MicrOMEGAs} \cite{Belanger:2001fz,Belanger:2004yn,Belanger:2006is,Alguero:2023zol}, or {\tt MadDM} \cite{Backovic:2013dpa,Backovic:2015cra,Ambrogi:2018jqj}.\footnote{There are many more DM tools publicly available, cf.~e.g.~the review \cite{Arina:2020alg}.} These tools calculate the relic density obtained from the freeze-out of WIMPs. The codes {\tt SuperIso Relic} and {\tt DarkSUSY} have (originally) been written for supersymmetric (SUSY) extensions. In the case of {\tt DarkSUSY}, the users can also directly provide the thermally averaged cross section (TAC) for their model to compute the relic density. In {\tt MicrOMEGAs} and {\tt MadDM} all matrix elements are automatically calculated during the execution of the program by {\tt CalcHEP} \cite{Belyaev:2012qa} and {\tt MadGraph5\_aMC@NLO} \cite{Alwall:2011uj}, respectively. They can also handle multi-component DM. All these codes rely on leading-order matrix elements. In {\tt MicrOMEGAs}, the TACs can be improved to include next-to-leading order (NLO) terms by editing predefined functions. For DM in the minimal supersymmetric extension (MSSM), the code {\tt DM@NLO} \cite{Herrmann:2007ku,Harz:2023llw} exists, which provides the NLO (SUSY) QCD calculation to DM  (co-)annihilation.
%and also elastic DM-nucleon scatter amplitudes for the %direct detection of DM. 
The programs furthermore compute the direct detection cross sections. For further details and codes including indirect DM searches, we refer to \cite{Arina:2020alg}. \s

In this paper, we present a new tool, the {\tt C++}~code \RE. The code computes the relic density based on the freeze-out mechanism for arbitrary models featuring a DM candidate stabilized by a discrete $\mathbb{Z}_2$ symmetry. The user solely has to provide the \texttt{FeynRules}~\cite{Alloul:2013bka} model files, and the relic density is then automatically computed by \RE. In this first release, the relic density is computed to leading-order in the thermally averaged cross section of the DM (co-)annihilation channels. The distinctly new feature of the code w.r.t.~existing codes is the scan procedure. The code implements an efficient scan algorithm that searches for parameter configurations that generate the correct relic density within the experimentally allowed limits, provided the model under investigation is able to saturate the relic density.  The code thereby provides a powerful tool to further constrain DM models by requiring the reproduction of the correct DM relic density. This can be used to subsequently investigate the phenomenological implications of this constraint for current and future collider and DM experiments on the one hand. On the other hand, this kind of investigation gives guidelines for further model building, in case the favored DM fails to reproduce the measured DM relic density.\s

In summary, the new DM code \RE\ includes significantly new features beyond the existing DM tools, which are:  
\begin{itemize}
\item The incorporation of efficient scan algorithms for parameter configurations within arbitrary DM models, that feature the correct relic density within the experimentally allowed uncertainties. The user solely needs to provide the model files and the code automatically calculates the DM relic density. 
\item The code has been set up such that it is extendable to include next-to-leading-order (NLO) corrections to the DM (co-)annihilation processes. The computation and validation of the NLO relic density in a specific DM model is in progress and will be published after this first release.
\item Having written the code ourselves, allows us to consistently combine it with the other codes generated by our collaboration to test new physics models with respect to particle, astroparticle and cosmological constraints, in particular our scanning tool {\tt ScannerS} \cite{Coimbra:2013qq,Muhlleitner:2020wwk}, which checks for theoretical and experimental constraints of extended Higgs sector models, and the code {\tt BSMPT} \cite{Basler:2018cwe,Basler:2020nrq,Basler:2024aaf}, which investigates BSM Higgs sectors with respect to their phase history in the early universe and possibly related gravitational waves signals. 
\end{itemize}

At present, the following beyond-SM models featuring DM candidates have been pre-imp\-le\-men\-ted and tested: the complex singlet extension of the Standard Model (CxSM) \cite{Coimbra:2013qq,Costa:2015llh,Barger:2008jx,Gonderinger:2012rd,Muhlleitner:2017dkd,Chiang:2017nmu,Egle:2022wmq,Egle:2023pbm}, 
%the Inert Doublet Model \cite{Deshpande:1977rw,Barbieri:2006dq,Dolle:2009fn,Diaz:2015pyv}
 the Next-to-2-Higgs-Doublet Model (N2HDM) in its dark doublet phase (DDP) \cite{Engeln:2020fld,Azevedo:2021ylf}, the model CP in the Dark (CPVDM) \cite{Azevedo:2018fmj,Biermann:2022meg,Biermann:2023owb}, the Two-Real-scalar-Singlet Model (TRSM) \cite{Ghorbani:2014gka,Robens:2019kga}, and the BDM5 \cite{Huang:2020ris, Capucha:2022kwo}.
New models, as stated above, can be included by providing the corresponding \texttt{FeynRules} model files. Note that the code at present does not handle models with more than one DM candidate. Furthermore, models where DM annihilation processes and the corresponding back-reactions differ (due to CP violation e.g.) are not supported either. 
Such extensions will be included in future releases. \s

The code can be downloaded from the url: 
\begin{center} 
\url{https://github.com/jplotnikov99/RelExt}
\end{center}

The outline of our paper is as follows. In Sec.~\ref{sec:codeoverview}, we give an overview of the code structure, followed by a description of the program in Sec.~\ref{sec:programdescription}. Section~\ref{sec:implementation} explains how a new model can be implemented by the users. Section~\ref{sec:algorithms} contains the detailed description of the algorithms used in \RE. In Sec.~\ref{sec:validation}, we present the validation of our code through the comparison with \texttt{MicrOMEGAs} for sample parameter points. We finish in Sec.~\ref{sec:conclusions} with the conclusions and the description of the next steps. The Appendix contains a comprehensive selection of sample results for the already implemented models in App.~\ref{app:examples}, a list of useful hints and advices for the generation of model files needed in the implementation of new models in App.~\ref{app:issues}, and in App.~\ref{app:mathematica} an overview of the \texttt{Mathematica} codes that, among other things, generate the relevant amplitudes squared for freeze-out.

\section{Code Overview \label{sec:codeoverview}}
\begin{figure}
    \centering
    \includegraphics[width=0.7\linewidth]{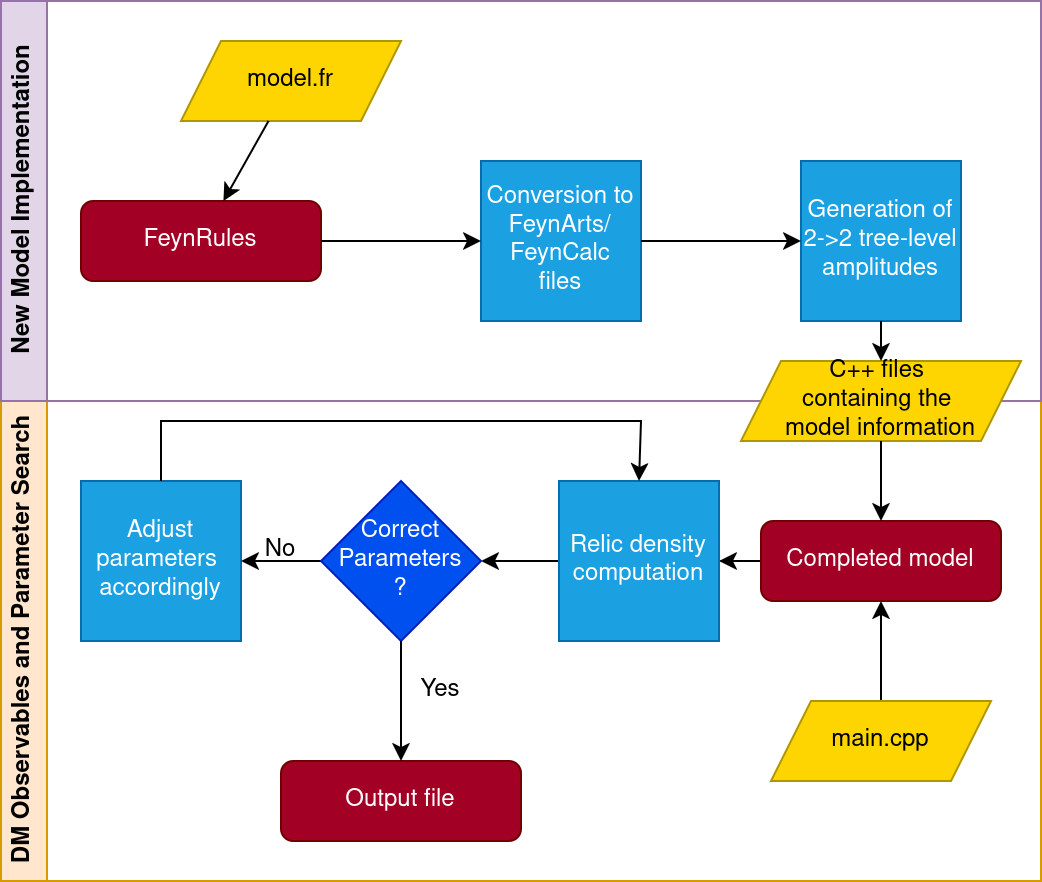}
    \caption{Flowchart of the code.}
    \label{fig:flowchart}
\end{figure}

We start by giving an overview of the code structure. The various building blocks will be explained in detail in the subsequent chapters. As visible from the flowchart depicted in Fig.~\ref{fig:flowchart}, the code consists of two main parts. They are given by the part for the "New Model Implementation" and the part for the "DM Observables and Parameter Search".\s

%Gelb: extern eingefügte Sachen; rot: Endstufen oder Anfangsstufen; hellblau: Zwischenstufen; dunkelblau: Abfragen

\paragraph{New Model Implementation}
This first part of the code allows the user to implement new models. This is done by providing the \texttt{FeynRules}~\cite{Alloul:2013bka} files for the model to be implemented. The code uses the packages \texttt{FeynArts}~\cite{Hahn:2000kx} and \texttt{FeynCalc}~\cite{Shtabovenko:2020gxv, Shtabovenko:2023idz} to automatically generate all tree-level DM annihilation amplitudes relevant for the calculation of the DM relic density. In this first release of the code these are the squared amplitudes for all tree-level DM $2\to2$ (co-)annihilation processes. These amplitudes will be saved together with additional necessary information about the model in \texttt{C++} files, which can then be compiled and used by the second part of the code. 
In case no new model is provided, the second part of the code can still be used with the already implemented models that are shipped with the code. 
The present version of the code is able to handle models with one $\mathbb{Z}_2$ symmetry, i.e.~with one DM candidate.
\paragraph{DM Observables and Parameter Search}
The second part of the code does not require \ttt{Mathematica}~\cite{Mathematica} and is purely written in \ttt{C++}. It performs the scans in the parameter spaces of the newly generated or already implemented models which search for parameter configurations that saturate the experimental limits on the relic density or that at least do not exceed it, in case weaker conditions are required. 
For this, the necessary thermally averaged cross sections are computed. This comprises also the computation of the needed total widths required for the virtual particle exchanges, and of the running couplings entering the interaction vertices of the scattering amplitudes. The code is designed such that it can be run with different settings and functions via a \ttt{main.cpp} file found within each model folder. The \ttt{main.cpp} allows the user to perform searches in the parameter space of a model using the provided functions (for more details see Sec.~\ref{sec: prog_use} and App.~\ref{app:examples}). The values of the calculated relic densities and the corresponding relevant input parameters are saved in the output file specified by the user.
\section{Program Description \label{sec:programdescription}}
In this section, we specify the system requirements for \DT, where to download the code, how to install it, and subsequently describe the usage of the program.

\subsection{System Requirements}
The program has been developed and tested on {\tt openSUSE Leap 15.4}, {\tt Ubuntu 22.04} and {\tt MacOS 13} with \ttt{g++ v.11}.  In order to run the code with all its features, $\DT$ needs the following programs to be pre-installed on the system:
\begin{itemize}
    \item \ttt{Cmake v3.22} or higher. It can be installed either through \ttt{pip} or directly from \url{https://cmake.org/download/}. It is used to compile $\DT$.
    \item \ttt{Mathematica v12}\cite{Mathematica} or higher is required for the implementation of new models. For the usage of implemented models, \ttt{Mathematica} and the packages listed here below are optional. 
    \item The \ttt{Mathematica} package \ttt{FeynRules}~\cite{Alloul:2013bka} is used to compute the Feynman rules for new models and to generate the required input files, which will be used by our code to implement new models.
    \item \ttt{FeynArts v3.11}~\cite{Hahn:2000kx} and \ttt{FeynCalc v9.3.1}~\cite{Shtabovenko:2020gxv, Shtabovenko:2023idz} are used to calculate the squared amplitudes of the DM annihilation processes. 
     
\end{itemize}

\subsection{Download}

The latest stable version of $\DT$ is available at 
\begin{center}
\url{https://github.com/jplotnikov99/RelExt}8
\end{center}
The user has the choice to clone the repository or download the program as a zip archive. The repository contains the following directories and files:

\begin{labeling}{mathematica}
    \item[\ttt{base\_Model}] This directory contains the necessary folders and files for each new model that is generated. It is copied and renamed when creating a new model.
    \item[\ttt{dataInput}] This directory is used to store the input files for the relic density calculation.
    \item[\ttt{dataOutput}] This directory is used to store the output files for the relic density calculation.
    \item[\ttt{include}] This directory contains the header files of the code.
    \item[\ttt{mathematica}] This directory contains the \ttt{Mathematica} files needed to implement a new model.
    \item[\ttt{md\_cpvdm}] This directory contains the already implemented model CPVDM~\cite{Azevedo:2018fmj,Biermann:2023owb}.
    \item[\ttt{md\_cxsm}] This directory contains the already implemented model CxSM~\cite{Coimbra:2013qq,Costa:2015llh,Barger:2008jx,Gonderinger:2012rd,Muhlleitner:2017dkd,Chiang:2017nmu,Egle:2022wmq,Egle:2023pbm}. 
     \item[\ttt{md\_DDP}] This directory contains the already implemented model N2HDM in its dark doublet phase~\cite{Engeln:2020fld,Azevedo:2021ylf}. 
      \item[\ttt{md\_trsm}] This directory contains the already implemented model TRSM~\cite{Ghorbani:2014gka,Robens:2019kga}. 
    \item[\ttt{md\_bdm5}] This directory contains the already implemented model BDM5 \cite{Huang:2020ris, Capucha:2022kwo}.
    \item[\ttt{numdata}] This directory contains the numerical data for the computation of the effective degrees of freedom \cite{GONDOLO1991145}.
    \item[\ttt{sources}] This directory contains the source files of the code.
    \item[\ttt{model}] Executable file to implement a new model.
    \item[\ttt{README.md}] Instructions file on how to run the code.
\end{labeling}

\subsection{Installation}
In the following description, we will refer to the directory chosen by the user for the installation of \DT, as \DTpath. In order to compile $\DT$ a \ttt{C++} compiler supporting the \ttt{C++14} standard is required. For the installation of \DT, the user has to do the following steps:
\begin{enumerate}
    \item In the directory \DTpath \,call
\begin{lstlisting}[language=bash]
 mkdir build && cd build
\end{lstlisting}
    \item  To generate the makefile, go to \DTpath\ttt{/build} and call
\begin{lstlisting}[language=bash]
 cmake [OPTIONS] ..
\end{lstlisting}
    with the following option
    \begin{itemize}
        \item $\ttt{CXX}$=$\ttt{C++Compiler}$. The default compiler is used if not provided.
    \end{itemize}
\item The last step is to call
\begin{lstlisting}[language=bash]
 make [ModelName] 
\end{lstlisting}
This will generate an executable in the directory $\DTpath$/\ttt{build} for a specific model that is implemented. If no models are provided when performing {\tt make}, all implemented models will be compiled. In the present release, these are the models CPVDM, CxSM, DDP, TRSM and BDM5, for which executables are generated. 
\end{enumerate}

\subsection{Program Usage}\label{sec: prog_use}
To run the executable for a given model, call from the directory $\DTpath$/\ttt{build}
\begin{lstlisting}[language=bash]
 ./[ModelName] [InputFile] [OutputFile]
\end{lstlisting}
This call will execute the \ttt{main.cpp} file for the corresponding model. With \ttt{InputFile} the user provides the input parameters of the model. The results computed in \ttt{main.cpp} are saved in the \ttt{OutputFile}. Every existing and newly generated model will have a \ttt{main.cpp} file which the users can modify for their own purposes. 
Each \ttt{main.cpp} consists of two sub-blocks which the user can modify. They are given by:

\begin{itemize}
    \item \textbf{Settings}: Here the user can set different options which are applied throughout the whole computation. 
    \item \textbf{Main}: This block is just the standard \ttt{main()} function of the \ttt{C++} language. Here, the user can call different functions provided by the code. To call these functions one has to create the object \ttt{Main}, which contains all of these functions. For newly generated models, this is done automatically. To call the functions simply type \ttt{M.[function name](args)}.
\end{itemize}
After making changes to the \ttt{main.cpp} file, call again in \DTpath\ttt{/build} 
\begin{lstlisting}[language=bash]
 make [ModelName] 
\end{lstlisting}
to apply the changes and call the corresponding executable to run the code.\s

All settings and functions are described in Tabs.~\ref{tab:standardsettings} and~\ref{tab:operations}, respectively.  Examples of input and \ttt{main.cpp} files for each \ttt{Mode} option, which determines the way parameter points are generated (see Tab.~\ref{tab:standardsettings}), can be found in the \ttt{dataInput} folder and the corresponding \ttt{md\_model} folder. To illustrate how a \ttt{main.cpp} file is set up to efficiently search for specific parameter regions (see Sec.~\ref{sec:paramSearches}) we provide in App.~\ref{app:examples} four examples for the three possible \ttt{Mode} options. These examples are included when downloading the code. \s

%
%%%%%%%%%%%%%%%%%%%%%%%%%%%%%%%%%
\renewcommand{\arraystretch}{1.1}
\begin{table}
\begin{tabular}
{p{0.24\textwidth}|p{0.7\textwidth}}
\hline
\ttt{Setting} & Description \\
\hline \hline
    \ttt{MODE} & Determines the way the parameter points are generated when calling the function \ttt{LoadParameters}. Currently, there are three possible options, which can be called by setting this option to 1, 2 or 3, respectively. In the first mode 1, only the initial parameter point is read in via the \ttt{InputFile}, which can then be altered in the \ttt{main} function. The user has to provide boundaries for each parameter in the \ttt{InputFile} to ensure reasonable convergence of the search algorithms. Mode 2 has the same structure for the \ttt{InputFile} as mode 1, but instead generates a random parameter point between the given parameter boundaries. For mode 3 the user provides a file with already generated parameter points, which are successively read in. The columns in this file must be tab separated.\\
    \ttt{SAVEPARS} & Determines which model parameters should be saved in the output file when calling the function \ttt{SaveData}.\\
    \ttt{CONSIDERCHANNELS}& Sets which (co-)annihilation channels need to be considered when calculating the relic density. Only these channels will contribute to the full relic density. The channels need to be provided in the format \ttt{"X1,X2,Y1,Y2"}, where Xi/Yi refer to the initial/ final state particles. To check which channels contribute to the relic density computation the user can call the function \ttt{PrintChannels}. If no arguments are provided all channels will be considered.\\
    \ttt{NEGLECTCHANNELS}&Sets which channels are to be ignored when computing the relic density. All channels except the ones provided here will contribute to the full relic density.\\
    \ttt{NEGLECTPARTICLES}
    &Sets which particles are to be neglected during the computation of the relic density. All channels that contain these particles will be ignored in the computation. To see the particle content of the model the user can call \ttt{PrintParticles}.\\
    \ttt{BEPS}& Sets the value of $B_\epsilon$ in Eq.~(\ref{eq: beps}) and Eq.~(\ref{eq: beps2})~(see below), which determines if (co-)annihilation channels are neglected and if the resonance peaks are integrated separately.\\
    \ttt{XTODAY} & Sets the value of $x_0$ in Eq.~(\ref{eq: fo-appr})~(see below). We recommend a value of $x_0=10^6$. A larger value would change the result only at the per mille level and may cause numerical instability for some points.\\
    \ttt{FAST}& If set to \ttt{true}, the freeze-out approximation given by Eq.~\eqref{eq: fo-appr}~(see below) is used when computing the relic density. Otherwise, the Boltzmann equation in \eqref{eq: boltz eq} is solved numerically. \\
    \ttt{CALCWIDTHS} & If set to \ttt{true}, the mediator decay widths will be computed by the code as described in Sec.~\ref{decaywidths}. If the users want to provide their own widths while using mode 3, this option must be set to \ttt{false}.\\
    \ttt{SAVECONTRIBS}& If set to \ttt{true} the function call \ttt{SaveData} will also save the contribution of each (co-)annihilation channel to the full relic density.\\
\hline
\end{tabular}    
\caption{Description of the options that can be used in the settings block of the \ttt{main.cpp} file.} 
\label{tab:standardsettings}
\end{table}
%%%%%%%%%%%%%%%%%%%%%%%%%%%%%%%%%%
\renewcommand{\arraystretch}{1.1}
\begin{table}[]
\begin{tabular}{p{0.24\textwidth}|p{0.7\textwidth}}
\hline
{\tt Function} & Description \\
\hline\hline
    \ttt{LoadParameters()}&Loads the external model parameters provided by the user in the input file. In mode 1 it loads the provided parameter point. In mode 2 it generates a random parameter point between the specified boundaries. In mode 3 it loads successively the next rows of the provided parameter points. \\
    \ttt{GetParameter(par)}&Returns the current value of the model parameter \ttt{par}.\\
    \ttt{ChangeParameter(par, val)}&Changes the current value of the model parameter \ttt{par} to \ttt{val}.\\
    \ttt{PrintParticles()}&Prints the particle content of the model.\\
    \ttt{PrintChannels()}&Prints all possible dark sector (co-)annihilation channels into particles that do not belong to the dark sector.\\
    \ttt{CalcXsec(smin, smax, Np, outfile, channel)}& Calculates the cross section (in units of GeV$^{-2}$) \ttt{Np} times for evenly distributed $\sqrt{s}$-values from \ttt{smin} to \ttt{smax} for a given \ttt{channel}. The $\sqrt{s}$-values with the corresponding cross sections are saved in \ttt{outfile} in the \ttt{dataOutput} folder. The format for \ttt{channel} is the same as in \ttt{CONSIDERCHANNELS}, i.e. \ttt{\{"X1,X2,Y1,Y2"}\}.\\
    \ttt{CalcTac(xmin, xmax, Np, outfile, channels)}& Computes the thermally averaged cross section (TAC) (in units of GeV$^{-2}$) \ttt{Np} times for evenly distributed $x$-values from \ttt{xmin} to \ttt{xmax}. The $x$-values and the corresponding TACs are saved in \ttt{outfile} in the \ttt{dataOutput} folder. Only (co-)annihilation channels provided in \ttt{channels} are considered. If no argument for \ttt{channels} is provided, \ttt{CalcTac} considers all possible (co-)annihilation channels.\\
    \ttt{CalcRelic()} & Returns the relic density via freeze-out for the currently set model parameters. This function will return 0 for unphysical parameters or if the freeze-out point cannot be found.\\\ttt{FindParameter(par, target, eps)} & Tries to find the value of the parameter \ttt{par} such that the relic density given by \ttt{target} is obtained within the provided uncertainty \ttt{eps}, i.e. $|\Omega h^2 - \Omega h^2_\text{target}|<\ttt{eps}$. The corresponding algorithm is described in Sec.~\ref{sec:paramSearches}.\\
    \ttt{RWalk(target, eps, gam, maxit)} & Does a random walk with the parameters given in the \ttt{InputFile} to obtain the \texttt{target} value of the relic density within the provided uncertainty \ttt{eps}. The parameter \ttt{gam} specifies the $\gamma_\text{max}$ described in Sec. \ref{sec:paramSearches} and \ttt{maxit} the maximum number of iterations. \\
    \ttt{InitMonteCarlo(Nbi, Nbe, pr, target)} & Initializes a grid on the parameters given in the input file. The region between each of the parameter boundaries is divided into \ttt{Nbi} bins of equal size. \ttt{Nbe} specifies how many cells will be tracked via the criterion given in Sec.~\ref{sec:paramSearches}. The variable \ttt{pr} specifies the probability with which a random point is generated. Lastly, \ttt{target} refers to the relic density the user wants to obtain.\\
    \ttt{SetWeight()} & Calling this function will compute the weight according to Eq.~(\ref{eq:monte weight}) with the last computed relic density and then adjust the best cells accordingly.\\
    \ttt{SaveData(pars)} & Saves the last computed relic density and the parameters given in the settings block into the output file.\\
    \hline
    \end{tabular}
\vspace*{0cm}    
\caption{Description of the functions that can be used in the main block of the \ttt{main.cpp} file.}    
    \label{tab:operations}
\end{table}
\section{How to Implement a new Model \label{sec:implementation}}

This section will provide a description on how to implement a new model in $\DT$. 
\begin{enumerate}
    \item To create a new model folder call
\begin{lstlisting}[language=bash]
 ./model -n [ModelName]
\end{lstlisting}
from $\DTpath$. This will create a folder named \texttt{md\_[ModelName]} in the $\DTpath$ directory. 
\item After creating the model folder, the \texttt{FeynRules} model files need to be stored in\\ \DTpath\texttt{/md\_[ModelName]/FR\_modfiles}. Afterwards call in $\DTpath$
\begin{lstlisting}[language=bash]
 ./model -l [path\to\FeynRules] [ModelFile.fr] [LagrangianName]
\end{lstlisting}
The first input parameter is the path to the $\texttt{FeynRules}$ directory. The second one is the name of the $\texttt{FeynRules}$ model file, while the last input is the name of the Lagrangian variable, which is defined in the model file. This command will automatically search the model file in $\DTpath$ and generate all relevant (co-)annihilation amplitudes for DM freeze-out. The generated \texttt{md\_[ModelName]} directory contains the following files and folders:
\begin{labeling}{\texttt{FR\_modfiles}}
    \item[\texttt{FR\_modfiles}] In this folder the model files generated by \texttt{FeynRules} are stored. The user has to save the \texttt{FeynRules} model files in this folder.
    \item[\texttt{sources}] This directory contains all the files that are necessary to perform the different operations to calculate the relic density within a given model, such as
    loading the various parameters, tokens and functions, calculating the widths and the running masses.
    \item[\ttt{sources/conditions.cpp}] This file contains the conditions that have to be fulfilled throughout a parameter scan. These conditions have to be provided by the user. An example is given below in App.~\ref{App: CPVDM}.
    \item[\texttt{sources/amp2s}] This directory contains the files with the squared amplitudes for all tree-level DM $2 \to 2$ (co-)annihilation processes, and the partial width(s) for the relevant mediator particle(s).
    \item[\texttt{model.hpp}] This file contains the declarations of all the parameters and functions of the model required for the relic density calculation.
    \item[\texttt{main.cpp}] This is the main file. The users can change this file using the provided functions to perform their own relic density parameter search.
\end{labeling}
\item The last step is to call in $\DTpath/\texttt{build}$
\begin{lstlisting}
 cmake..
 make [ModelName]
\end{lstlisting}
This command will create an executable for the implemented model.\\
To delete a model do not simply delete the corresponding folder, but instead call in $\DTpath$
\begin{lstlisting}
 ./model -d [ModelName]
\end{lstlisting}
\end{enumerate}
\section{Structure and Description of the Algorithms \label{sec:algorithms}}
In this section we will describe the numerical methods and algorithms that are used throughout the program.\s

The BSM models that provide DM candidates depend on a set of input parameters that can be more or less lengthy. Depending on the model and the interplay of the parameter values and the constraints on the models, more or less extensive parameter scans have to be performed, in order to find viable relic density values and to perform meaningful phenomenological investigations. A main objective of the code is therefore to solve the Boltzmann equation given in terms of the yield $Y$ by \cite{PhysRevD.56.1879}
\begin{equation}\label{eq: boltz eq}
 \frac{\mathrm{d}Y}{\mathrm{d}x} = - \sqrt{\frac{\pi}{45G}}\frac{g^{1/2}_*}{x^2}\langle\sigma v\rangle_\text{eff} \left(Y^2 - Y_\text{eq}^2\right) \, ,
\end{equation}
as fast and as precisely as possible. Here, $x=m_1/T$ is the DM mass $m_1$ divided by the temperature $T$. The gravitational constant is denoted by $G$, and $g^{1/2}_*$ is given by 
\begin{equation}
    g^{1/2}_* = \frac{h_\text{eff}}{\sqrt{g_\text{eff}}}\left (1+\frac{T}{3h_\text{eff}}\frac{\mathrm{d}h_\text{eff}}{\mathrm{d}T}\right) \,,
\end{equation}
where $h_\text{eff}$ and $g_\text{eff}$ are the effective degrees of freedom of the entropy density and the energy density, respectively. The equilibrium yield $Y_\text{eq}$ in the non-relativistic limit is given by
\begin{equation}
    Y_\text{eq} = \sum_i Y_{i,\text{eq}} = \frac{45 x^2}{4\pi^4 h_\text{eff}(x)}\sum^N_{i} g_i \cdot\left(\frac{m_i}{m_1} \right)^2 \cdot K_2\left(\frac{m_i}{m_1}x\right) \,,
\end{equation}
where the sum is performed over all dark sector particles $i$, with $m_i$ denoting their corresponding mass. The TAC $\langle\sigma v\rangle_\text{eff} $ for the {(co-)annihilation} of the dark sector particles $i$ and $j$ into the thermal bath particles, is given by
\begin{equation}\label{eq: TAC}
    \langle\sigma v\rangle_\text{eff} = \frac{\sum_{i,j=1}^N g_i g_j \int_{(m_i+m_j)^2}^\infty \mathrm{d}s \sqrt{s}p_{ij}^2 \sigma_{ij}(s) K_1\left(\frac{\sqrt{s}x}{m_1}\right)}{2T\left(\sum_{i=1}^N g_i m_i^2K_2\left(\frac{m_i}{m_1}x\right)\right)^2} \, .
\end{equation}
The $K_n$ ($n=1,2$) are the modified Bessel functions of the second kind at order $n$, $g_i$ and $g_j$ are the internal degrees of freedom of the dark sector particles and $\sqrt{s}$ is the center-of-mass energy of the (co-)annihilation process. Furthermore, $p_{ij}$ is given by
\begin{equation}
    p_{ij} = \frac{\sqrt{(s-(m_i+m_j)^2)(s-(m_i-m_j)^2)}}{2\sqrt{s}} \,.
\end{equation} 
The cross section $\sigma_{ij}$ in Eq.~\eqref{eq: TAC} 
denotes the (co-)annihilation of two dark sector particles $i,j$ into all possible freeze-out final states, where in the present released version we only consider two-body final states $k$, $l$ such that
\begin{equation}\label{eq: xsec}
    \sigma_{ij} = \frac{1}{32\pi sg_ig_j} \sum_{k,l} \frac{\lambda^{1/2}(s,m_k^2,m_l^2)}{\lambda^{1/2}(s,m_i^2,m_j^2)}\int |\mathcal{M}_{ij\rightarrow kl}|^2 d\cos\theta \,. 
\end{equation}
Here, $\lambda$ is the Käll\'en function, $\theta$ the azimuthal scattering angle, and $\mathcal{M}_{ij\rightarrow kl}$ the $2\rightarrow2$ \mbox{(co-)annihilation} matrix element. After solving Eq.~\eqref{eq: boltz eq} from $x=0$ to $x=x_0$, the relic density can be calculated using \cite{BELANGER2007367}
\begin{equation}
    \Omega_\text{DM}h^2=2.742\cdot10^8\frac{m_1}{\text{GeV}}Y(x_0)\,,
\end{equation}
where $x_0=m_1/T_0$ and $T_0$ is the temperature of the universe today.

\subsection{Amplitude Generation \label{sec:ampl}}

To calculate the relic density, we first need to generate the squared (co-)annihilation matrix elements $|\mathcal{M}_{ij\rightarrow kl}|^2$. For their calculation, we use the \texttt{Mathematica} code described in App.~\ref{app:mathematica}, with the packages \texttt{FeynArts} and \texttt{FeynCalc}. The code recognizes automatically all dark sector particles from the given model file\footnote{Particles with a tilde before their name belong to the dark sector (see App. \ref{app:issues}).} and then computes all different (co-)annihilation channels into particles from the visible sector. To avoid unnecessary computation time, we tokenize the amplitudes squared. Here, couplings or combinations of couplings that appear more than once, are stored in tokens which will be only computed once per parameter point. To avoid possible divergences coming from propagators in the amplitudes, we use the Breit-Wigner formula, and make the replacement
\begin{equation}
\frac{1}{s-m_j^2} \rightarrow \frac{1}{s-m_j^2 - i\Gamma_{j} m_j } \, ,
\label{eq:propagator}
\end{equation}
where $m_j$ and $\Gamma_j$ are the mass and total decay width, respectively, of the mediator particle $j$.
The code automatically identifies all $s$-channel mediator particles and calculates their total decay widths. We will give an overview on how the decay widths are obtained in Sec.~\ref{decaywidths}. Additionally, in some kinematic regions it is possible to encounter $t/u$-channel divergencies. To regulate these divergencies we use a similar approach to {\tt MicrOMEGAs} and introduce a width via $\Gamma_j=m_j/1000$ \cite{Alguero:2023zol}. Using this width leads to a deviation of less than 1\% outside of the divergent regions compared to the zero width result.
%Since these only occur in rare instances we cure them by introducing a width which is 1\% of the mass of the mediator particle, i.e.
%\begin{equation}
%    \frac{1}{t-m_j^2} \rightarrow \frac{1}{t-m_j^2 - 0.01\cdot im_j^2 } \, ,
%\end{equation}
\\
For DM (co-)annihilation into quark pairs, we use running quark masses in the final states, unless for the top quark, where we use the pole mass value, which must be set by the user. The running quark masses are evaluated at next-to-next-to-next-to leading log accuracy (N$^3$LL) and taken at the scale given by twice the DM mass. We take as input values the following $\overline{\mbox{MS}}$ masses for the bottom ($b$), charm ($c$) and strange ($s$) quark, respectively, at the scale $Q$ given in the brackets: $\overline{m}_b (\overline{m}_b) = 4.18$~GeV, $\overline{m}_c (3\mbox{ GeV}) = 0.986$~GeV, and $\overline{m}_s (2\mbox{ GeV}) = 0.095$~GeV. At the implemented loop order, this corresponds to the following bottom and charm pole masses, $m_b^{\text{pole}}= 4.83$~GeV and $m_c^{\text{pole}}= 1.42$~GeV. The quark masses of the first two generations and the lepton masses must be provided by the user. In the computation of the total widths, running masses are used in the way described in the next section. Furthermore, when needed, we use the strong coupling constant at the scale given by twice the DM mass in the DM (co-)annihilation processes. In the computation of the total decay widths, we evaluate the strong coupling constant at the scale given by the mass of the decaying particle. The strong coupling constant is computed at next-to-next-to-next-to leading order (N$^3$LO). \s

The expressions of the squared amplitudes will subsequently be converted in \texttt{C++} format and stored. We note that no spin or color average factors are included in these squared amplitudes. They will be included when the (thermally averaged) cross sections are calculated.

\subsection{Computation of the Total Widths}\label{decaywidths}
The (co-)annihilation cross section $\sigma_{ij}$ proceeds via mediator particle(s), which can become on-shell for certain energy values $\sqrt{s}$. We therefore have to include the total widths in the propagators of the exchanged particles in order to avoid poles in the calculation of the TAC. As described in Sec.~\ref{sec:ampl}, these total widths are computed automatically (apart from the exceptions described hereafter) by the code as the sum of the automatically calculated partial decay widths of the mediator particles. Apart from the scalar/pseudoscalar mediators, all decay widths are calculated for $1 \to 2$ particle decays and at leading order. In this subsection, we give some further details on the implementation. \s

In the case of massive SM gauge bosons $V$ acting as mediators, where $V=W^\pm, Z$, the total widths are taken from the input model files, as they are measured to experimentally high precision. In the case of the top-quark acting as a mediator, the total decay width is computed automatically by \DT. This can lead to total top decay widths that do not comply with the experimentally measured value. The users themselves have to make sure to choose only such parameter configurations for their model that do not violate the experimental constraints. The total widths of all other mediator particles are obtained automatically by calculating and summing up their tree-level partial decay widths. \s

In the computation of the total widths for scalar and pseudoscalar mediator particles, we include the available dominant higher-order corrections to the decay widths, as explained in more detail in the following. First of all, for the  scalar/pseudoscalar mediator particles, analytical formulae are implemented in \DT\ and used for the calculation of the partial decay widths, which are summed up to the total widths. We follow here mostly the implementation of {\tt HDECAY} \cite{Djouadi:1997yw,Djouadi:2018xqq}, a code for the computation of the Higgs decays for the SM and the minimal supersymmetric SM extension (MSSM), including state-of-the-art higher-order corrections, and take the following formulae given in Ref.~\cite{Spira:2016ztx}, unless stated otherwise. Here, we include the coupling modification factors of the scalar/pseudoscalar mediator to the SM particles according to the model under investigation. For the decays into lepton final states, we take the tree-level formula given in Eq.~(21). 
Since the QCD interactions are not affected by modifications of the Higgs sector, we include higher-order QCD corrections in the decays into quarks. We interpolate here between the large Higgs mass region and the threshold region, and we include the relative QCD corrections $\delta_{\text{QCD}}$ and the relative top-quark induced contributions $\delta_t$, as given in~Eq.~(23). For the large Higgs mass region, we apply formula Eq.~(15) for $\delta_{\text{QCD}}$, using running $\overline{\mbox{MS}}$ quark masses both at tree-level and at higher order. These and the running strong coupling constant $\alpha_s$ with five active flavours are taken at the scale of the mass of the decaying Higgs, thereby absorbing large logarithms. The $\delta_t$ corrections differ for the scalar and pseudoscalar case and are given after Eq.~(23). In the threshold region, finite quark mass effects become relevant, and we apply Eq.~(25). The QCD corrections are given in Eq.~(17) for the scalar and in Eq.~(26) for the pseudoscalar case, where we now take quark pole masses everywhere, both at leading and at higher order. The coupling modification factor for the Higgs  coupling to the quarks in Eq.~(23) and Eq.~(25), respectively, is automatically replaced by the corresponding one for the model under consideration. \s

For the LO decays into gluons we use Eqs.~(61,62) for the scalar and Eqs.~(65,66) for the pseudoscalar case. The higher-order QCD corrections are taken into account by applying Eq.~(51) for the scalar and Eq.~(67) for the pseudoscalar case, for five active flavours and the strong coupling constant taken at the scale of the Higgs mass. In case the user included in the \ttt{FeynRules} model files effective couplings between the scalars/pseudoscalars and the gluons, these will not be used in the computation of the partial widths, which will not have any significant effect on the final relic density since these partial widths are very small. For the actual calculation of the relic density, however, the effective couplings provided by the user will be used. \s

Additionally, off-shell decays into massive gauge bosons $V$ ($V=W^\pm, Z$) are taken into account. In a CP-conserving theory, they are only allowed for the scalar mediators. For energies
$\sqrt{s} \ge 2 m_V$, we apply the on-shell decay formula, given by Eqs.~(34--36). Since the QCD corrections are of moderate size, we do not include them in our calculation. This also ensures a smooth transition to the off-shell decays, where no QCD corrections are available in a compact form. For energies below the production of any on-shell gauge bosons, $\sqrt{s} \le m_V$, we apply Eqs.~(37--39) for the production of two off-shell gauge bosons. In the energy range $m_V \le \sqrt{s} \le 2 m_V$, we compute the decay into one off-shell and one on-shell gauge boson, using Eqs.~(51,52), given in \cite{Spira:1997dg}. \s

In the present version, the loop-induced decays into photon pairs or a $Z$ plus photon pair are not considered for the computation of the total width, as these decay widths are very small and barely modify the total width. \s

In BSM models, further Higgs decays are possible, which, depending on the parameter region, can become sizeable, or even dominant and therefore have to be included in the total widths as well. These can be e.g.~decays into a lighter Higgs plus gauge bosons, lighter Higgs pairs, new heavy fermions or new heavy gauge boson pairs. The code computes all $1\rightarrow2$ non-SM-like decays and adds them to the decay widths into SM particles to obtain the total width.

\subsection{Computation of the Thermally Averaged Cross Section}

With the generated amplitudes we move on to the computation of the TAC given in Eq.~\eqref{eq: TAC}. We start by integrating Eq.~\eqref{eq: xsec} over $\cos\theta$. For this, we compute a first estimate of the integral using a composite Simpson's $3/8$ rule \cite{10.5555/1403886}. We use this estimate as part of the termination criterion for the adaptive Simpson's $3/8$ method with which we compute the cross section. The use of the estimate in the termination criterion ensures a rapid convergence in areas where the integrand is small. 
Next, we need to integrate the numerator of Eq.~\eqref{eq: TAC}. In order to do this we first sort the (co-)annihilation processes by  the magnitude of their rest masses via
\begin{equation}
    \text{max}\left(m_i+m_j, m_k+m_l\right)\; .
\end{equation}
This allows us to adjust the lower integration bound accordingly for each process. Additionally, we neglect channels which are heavily Boltzmann suppressed, i.e., that do not fulfill the criterion  \cite{Belanger:2006is}
\begin{equation}
    B=e^{-x\frac{m_{i(k)}+m_{j(l)} - 2m_1}{m_1}}>B_\epsilon\;,\label{eq: beps}
\end{equation}
with the recommended value $B_\epsilon=10^{-6}$. Further, we approximate the modified Bessel functions of the second kind via \cite{zbMATH03273551}\footnote{To ensure that the precision of our results is not influenced by this, we use this approximation only for $x>5$.}
\begin{equation}
    K_n(x)\approx\sqrt{\frac{\pi}{2x}}e^{-x}\left(1+\frac{4n^2-1}{1!(8x)}+\frac{\left(4n^2-1\right)\left(4n^2-9\right)}{2!(8x)^2}+...\right)\,.
\end{equation}
This approximation results in an increase of the numerical stability of the TAC for large values of $x$ and decreases the computational effort. Similar to the $\cos\theta$ integration, we compute a first estimate of the TAC integral using the Kronrod 61 point method \cite{10.5555/1403886} for the regions outside of $s$-channel resonance peaks. As in the $\cos\theta$ integration, we use this estimate in the termination criterion to compute the integral using an adaptive Gauss-Kronrod 15 point method \cite{10.5555/1403886}. Since resonance peaks are challenging to deal with numerically, we integrate them separately if they are close enough to the energy of the rest mass of the considered channel, i.e.~if the condition
\begin{equation}
    e^{-x\frac{m_\text{med}-m_{i(k)}+m_{j(l)}}{m_1}}>B_\epsilon\;,\label{eq: beps2}
\end{equation}
is fulfilled, with $m_\text{med}$ being the mass of the $s$-channel mediator particle.
In this case the peaks are properly taken into account by integrating them at high precision using the adaptive Simpson's $3/8$ method. \s

Note that for each $s$ value that we integrate over, we first need to integrate over $\cos\theta$ to obtain $\sigma_{ij}(s)$. Doing this for different $x$ values would amount to a large computational effort. However, we note that $\sigma_{ij}$ is independent of $x$. We therefore compute $\sigma_{ij}$ only once at different $s$ values and save these results in a hash map in order to avoid unnecessary redundant computation.

\subsection{Boltzmann Equation \label{sec:Beq}}

After the computation of the TAC, we can move on to the solution of Eq.~\eqref{eq: boltz eq}. For the effective degrees of freedom $g_\text{eff}$, $h_\text{eff}$ and $g_*^{1/2}$ we use interpolated values from \cite{GONDOLO1991145}. We determine the freeze-out point $x_\text{f}$ via the condition given in \cite{GONDOLO1991145}, 
\begin{equation}\label{eq: initial condition}
    \sqrt{\frac{\pi}{45G}}\frac{g_*^{1/2}m}{x^2}\langle\sigma v\rangle_\text{eff}Y_\text{eq}\delta(\delta+2)+\frac{\text{d}\ln Y_\text{eq}}{\text{d}x}=0\;,
\end{equation}
where $\delta$ can be freely chosen. To find the point $x_\text{f}$ which fulfills Eq.~\eqref{eq: initial condition}, we use a Bisection method\footnote{We only search between $x=5$ and $x=50$. If a point is not found within these boundaries it is not a valid freeze-out point and we return 0 for the relic density.} to find the root of the equation. If the option \texttt{Fast} is set to true we use $\delta=1.5$ as recommended in \cite{GONDOLO1991145} and compute the yield via the freeze-out approximation, i.e.
\begin{equation}
    \frac{1}{Y_0}=\frac{1}{2.5Y_\text{eq}(x_\text{f})}+\sqrt{\frac{\pi}{45G}}\int_{x_\text{f}}^{x_0}\text{d}x\frac{g_*^{1/2}m}{x^2}\langle\sigma v\rangle_\text{eff}\;,\label{eq: fo-appr}
\end{equation}
where the $Y_\text{eq}^2$ term in Eq.~\eqref{eq: boltz eq} was discarded. As stated in \cite{Belanger:2004yn} and \cite{GONDOLO1991145}, this approximation is accurate within a few percent in almost all scenarios, which we can confirm. When the \texttt{Fast} option is turned off we use $\delta=0.1$ to find the freeze-out point. With this we solve Eq.~\eqref{eq: boltz eq} numerically with the initial condition $Y_\text{f}=1.1Y_\text{eq}(x_\text{f})$ at the freeze-out point using an adaptive Dormand-Prince 853 method as described in \cite{10.5555/1403886}. 

\subsection{Parameter Searches \label{sec:paramSearches}}
Depending on the complexity of the provided model, finding parameter regions that acquire the desired relic density can vary in difficulty. For this reason, we currently provide three different methods to search for such regions. They are described in the following. \s

The {\it first method} uses a Monte Carlo approach by laying a grid over the allowed parameter space, consisting of $N_\text{b}^{N_\text{p}}$ cells, where $N_\text{b}$ is the number of bins per parameter and $N_\text{p}$ is the number of parameters we want to scan. To initialize this grid the user has to call the function \ttt{InitMonteCarlo} with the corresponding function parameters (see Tab.~\ref{tab:operations}). Each time a new relic density $\Omega_c$ is computed the user can call the \texttt{SetWeight} function which assigns a weight to the current parameter point. This weight depends on how far away the result $\Omega_c$ is from the desired value $\Omega_d$ and has a value $w$ which ranges from 0 to 1, given by
\begin{equation}
    w = \begin{cases}
        \left(\frac{\Omega_d}{\Omega_c}\right)^2,&\Omega_d<\Omega_c\\
        \left(\frac{\Omega_c}{\Omega_d}\right)^2,&\Omega_d>\Omega_c
    \end{cases}
    \;.
    \label{eq:monte weight}
\end{equation}
This means that the closer $w$ is to 1 the closer we are to the desired relic density and vice versa. The code keeps track of up to $N_\text{best}$ best cells with respect to $w$. If we find an $\Omega_c$ in a new cell which is better than the worst one in the tracked cells, that cell will get substituted by the new cell. However, if a new $\Omega_c$ with a larger $w$ is calculated in one of the tracked cells, that cell will be updated with the new $w$. These tracked cells also influence the generation of new parameter points. Once we have obtained $N_\text{best}$ cells we generate new parameter points from these cells with a probability $p_b$ and random points with a probability $p_r=1-p_b$. At the end of the scan we create an additional output file in which the parameter regions of these cells are stored and can be used for future scans. \s 

In the {\it second method}, which can be called via \texttt{FindParameter}, we only change a single parameter trying to obtain $\Omega_d$. The algorithm starts by going into the desired direction in large steps by following the gradient such that $\Delta\Omega=\Omega_c-\Omega_d$ gets minimized. 
There are now two possible scenarios. The first one is that at some point during this minimization procedure $\Delta\Omega$ and the gradient flip their sign. This means that the root of $\Delta\Omega$ has to be between the current parameter value and the last one. The code recognizes this and switches to a bisection method to find the root. The second scenario is one in which the sign of $\Delta\Omega$ stays the same, while the sign of the gradient flips. In such a case we have a local minimum. The code will then switch to a gradient descent method to find this local minimum. However, if during the descent method $\Delta\Omega$ flips its sign we switch to the bisection method to find the root. In short, this method either finds the closest root or minimum of $\Delta\Omega$ by only changing a single parameter. The routine \texttt{FindParameter} will return the last parameter considered and the corresponding relic density or the value at the minimum if it is unable to find a solution such that $|\Delta\Omega| <$ \ttt{eps}, where \ttt{eps} is the uncertainty set by the user. This can happen for a few different reasons, like e.g.: there is no solution, a minimum was found before the (actually existing) solution was reached, the solution is outside the boundaries specified by the user, or the solution could not be found before the limit of iterations was reached. In these cases, changing the initial point, the boundaries set for the parameters (see App.~\ref{app:examples}), the uncertainty for $\Omega_d$, and/or the parameters that are being changed, are some of the options to find a solution, depending on the model under investigation. \s

The {\it third method}, which can be used via the function \texttt{RWalk},  performs a random walk algorithm with multiple parameters. Here we change each parameter value $x_i$ to a new one $x_i'$ via 
\begin{equation}
    x_i' = x_i(1+\gamma_i)\;.\label{eq: rwalk}
\end{equation}
The $\gamma_i$ is randomly chosen from the range $\left[-\gamma_\text{max},\gamma_\text{max}\right]$, where $\gamma_\texttt{max}$ is by default set to $0.01$. We then compare the $\Delta\Omega$ computed with the parameters $x_i$ with the $\Delta\Omega$ computed with $x_i'$. If we see an increase in $\Delta\Omega$ we try a new set of points generated via Eq.~(\ref{eq: rwalk}). However, if we see a decrease, we continue applying the same set of $\gamma_i$'s until we see an increase in $\Delta\Omega$. The algorithm stops when the desired relic density is found within the specified uncertainty, or when the maximum iteration limit of the random walk is reached. 
\section{Validation \label{sec:validation}}

\begin{figure}[t!]
	\begin{center}
	\includegraphics[width = 10cm]{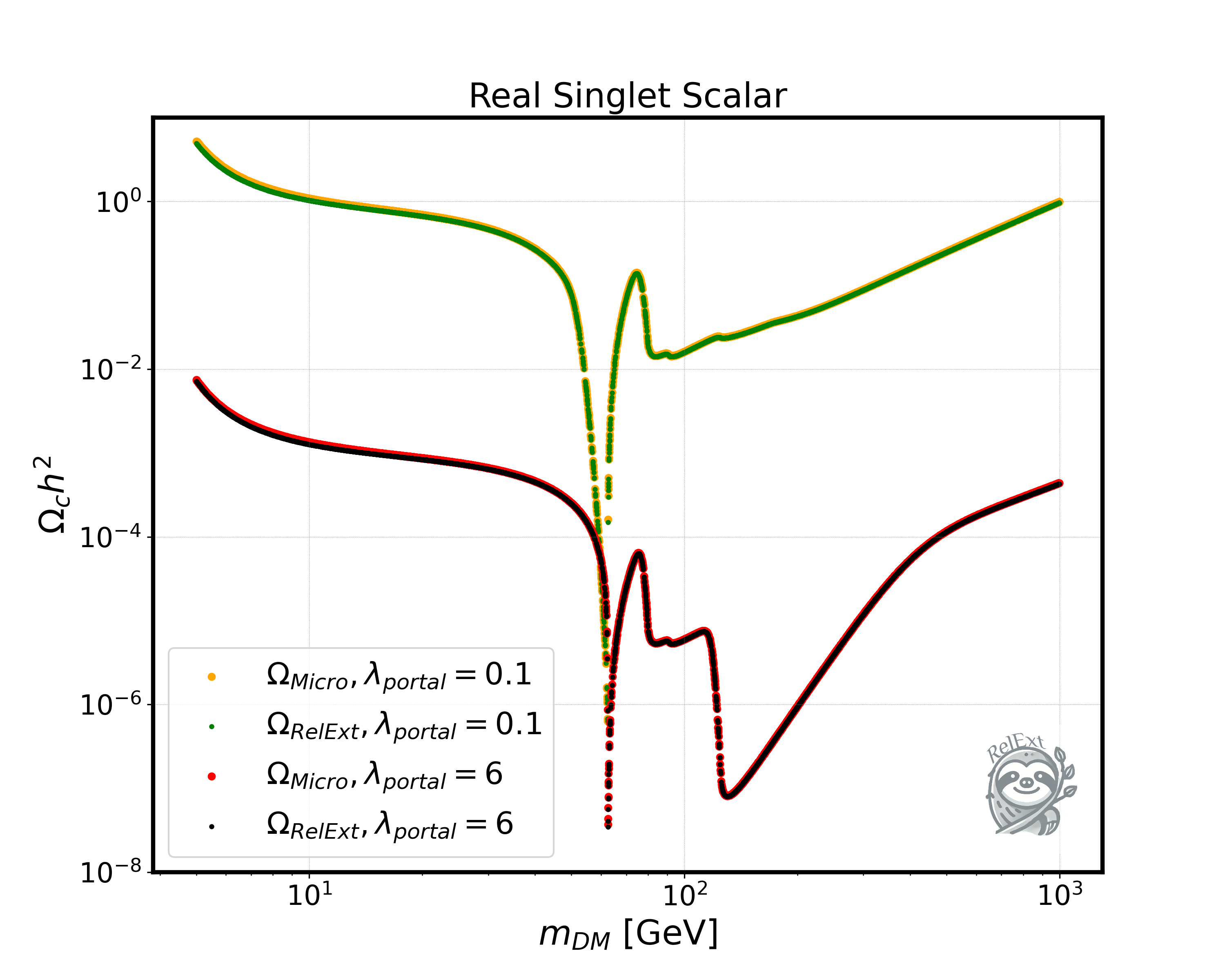}
	\caption{Relic density for the real singlet model, as a function of the DM mass computed with \texttt{RelExt} or \texttt{MicrOMEGAs}. For the orange and red points, we used \texttt{MicrOMEGAs} and set the Higgs portal coupling to 0.1 and 6, respectively. For the green and black points, we used \texttt{RelExt} and also set the Higgs portal coupling to 0.1 and 6, respectively.}
	\label{fig:comparison1}
	\end{center}
\end{figure}

\begin{figure}[t!]
	\begin{center}
	\begin{tabular}{c}
	\includegraphics[width = 10cm]{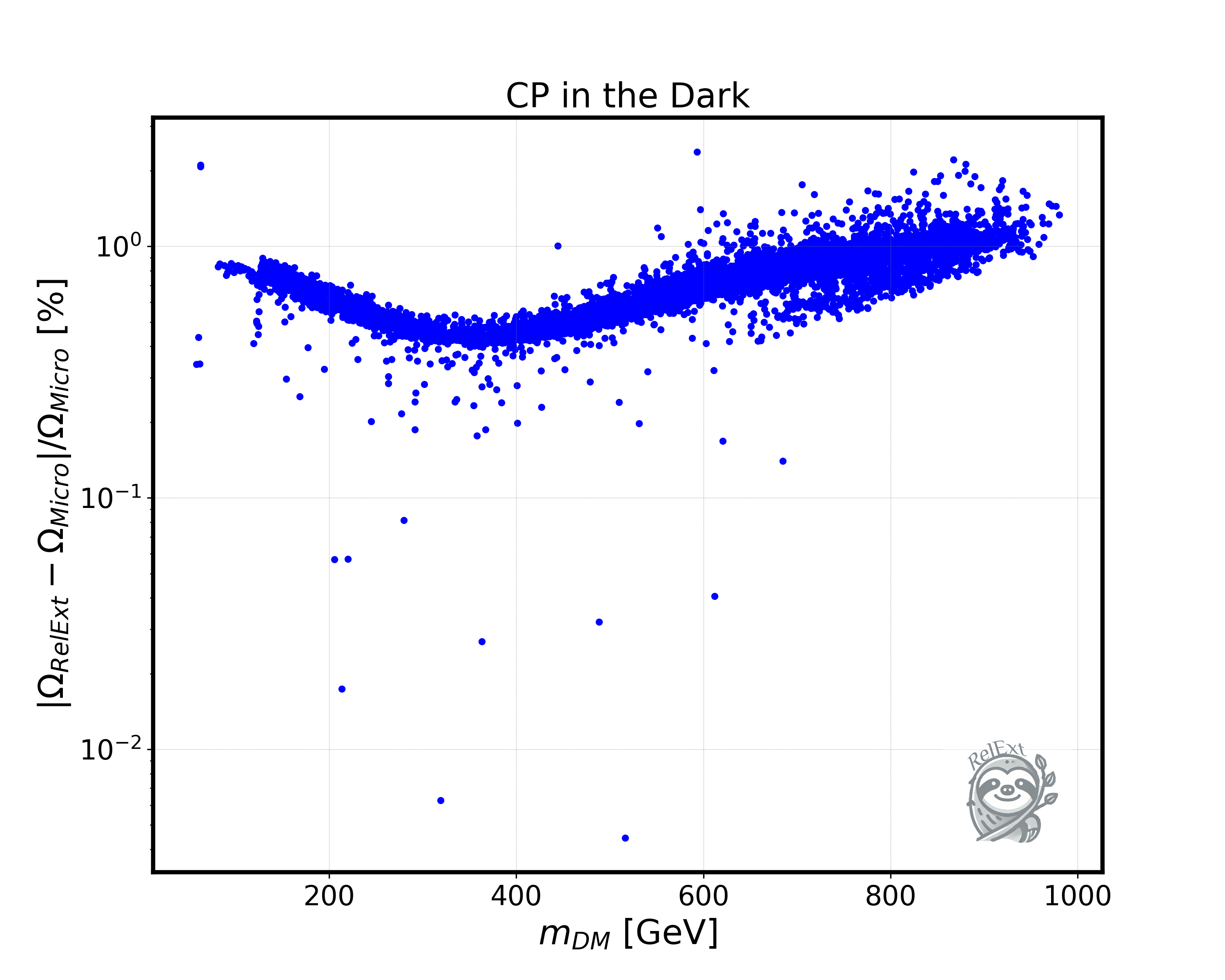}
	\end{tabular}
	\caption{Absolute value for the relative difference of the relic densities calculated with \texttt{RelExt} and \texttt{MicrOMEGAs}, respectively, as a function of the DM mass, for the model CP in the Dark.}
	\label{fig:comparison2}
	\end{center}
\end{figure}

\begin{figure}[h!]
	\begin{center}
	\begin{tabular}{c c}
    \includegraphics[width = 8cm]{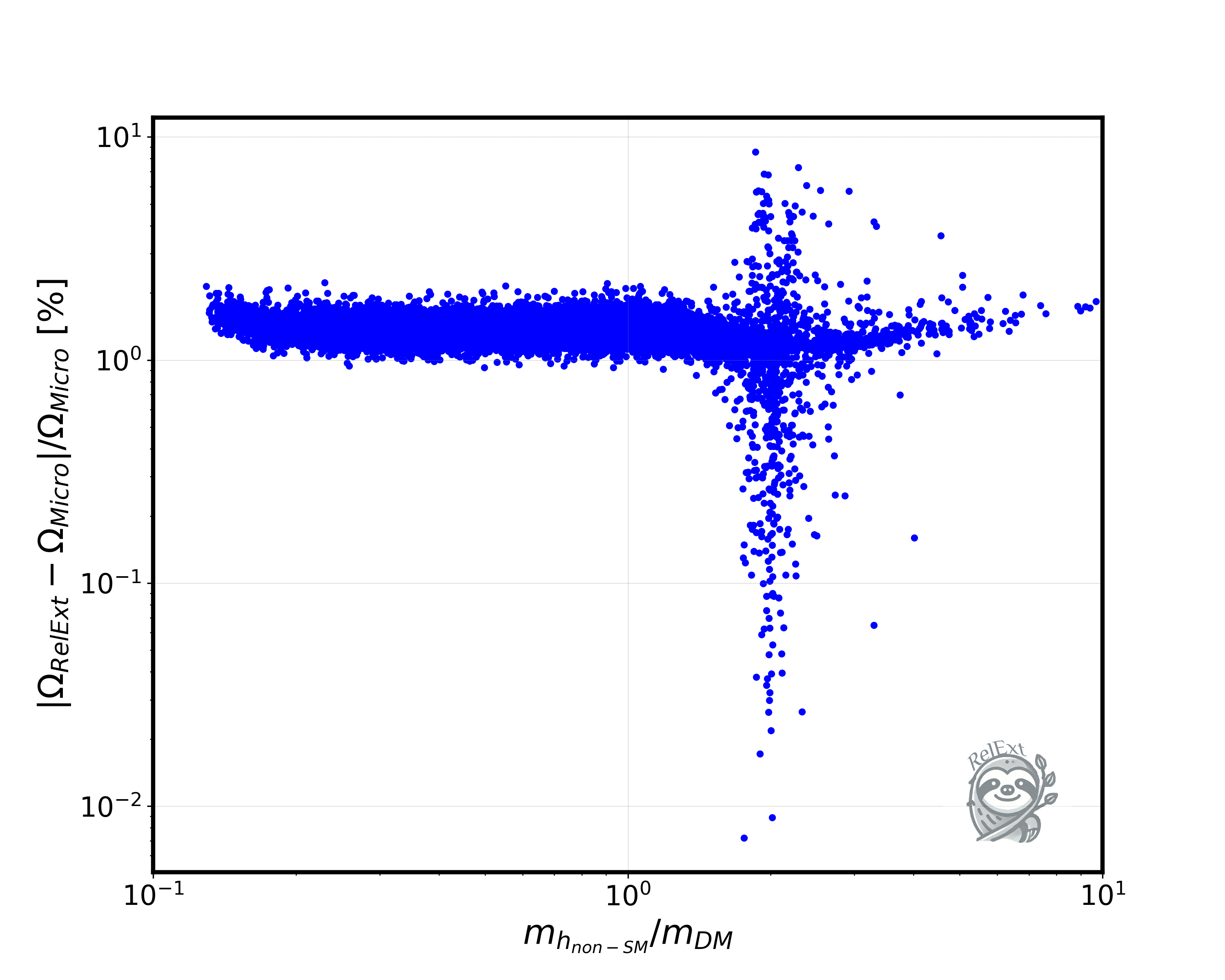} &
	\includegraphics[width = 7.9cm]{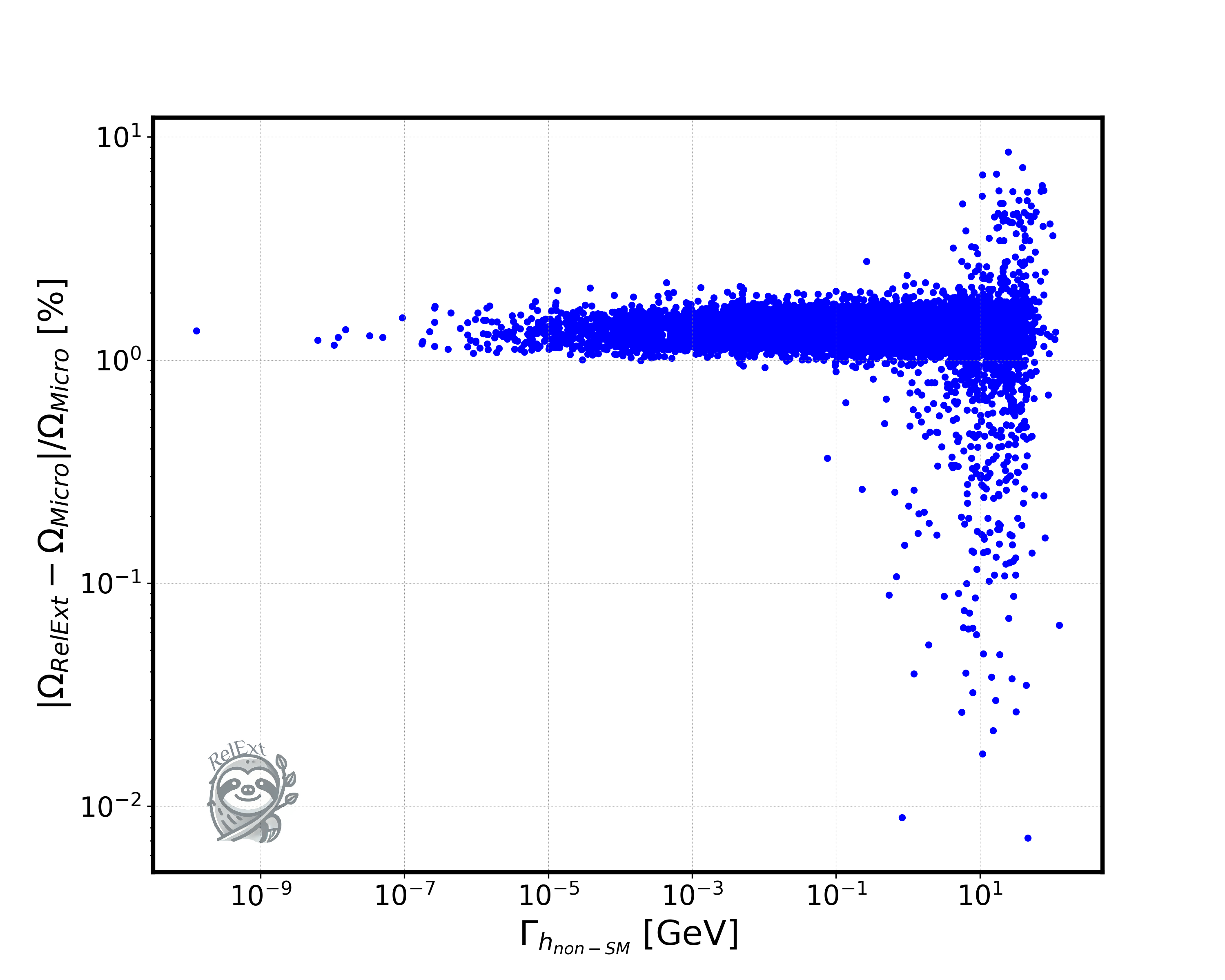}  
	\end{tabular}
	\caption{Absolute value for the relative difference of the relic densities calculated with \texttt{RelExt} and \texttt{MicrOMEGAs}, respectively, as a function of $m_{h_{\text{non-SM}}}/m_{DM}$ (left) and $\Gamma_{h_{\text{non-SM}}}$ (right), for the dark doublet phase of the N2HDM. Here, $m_{h_{\text{non-SM}}}$ is the mass of the additional non-SM-like Higgs boson in the visible sector, and $\Gamma_{h_{{\text{non-SM}}}}$ its width.}
	\label{fig:comparison_widths}
	\end{center}
\end{figure}

\begin{figure}[ht]
	\begin{center}
	\begin{tabular}{c}
	\includegraphics[width = 10cm]{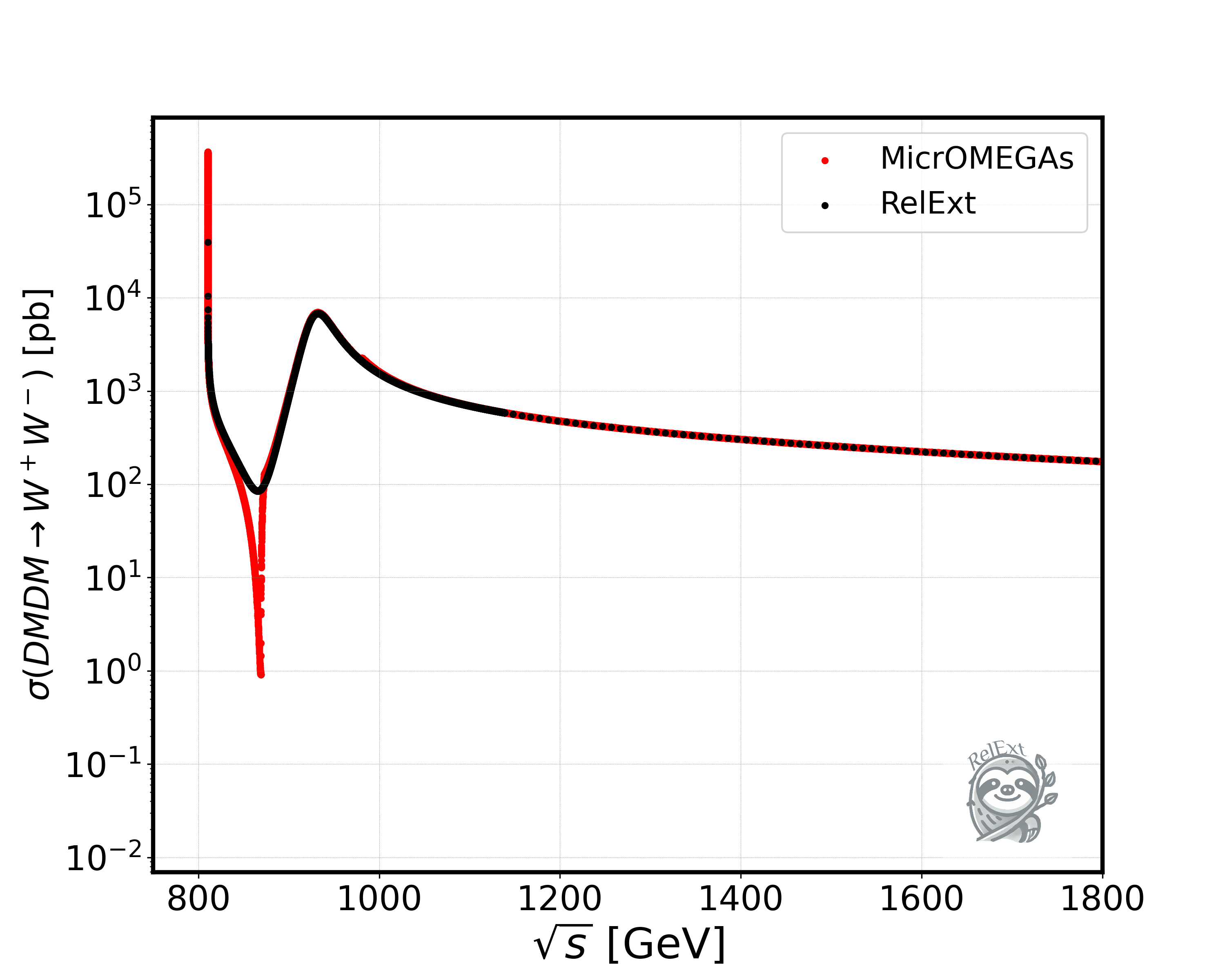}
	\end{tabular}
	\caption{Dark Matter annihilation cross section $\sigma(\text{DM DM} \to W^+W^-)$ in pb, as a function of the center-of-mass energy $\sqrt{s}$ in GeV, for the dark doublet phase of the N2HDM. The black points are obtained with \texttt{RelExt} and the red points with \texttt{MicrOMEGAs}.}
	\label{fig:comparison_xsec}
	\end{center}
\end{figure}

For each of the models implemented in the code, the relic density was computed ($\Omega_{\text{RelExt}}h^2$) and compared with the one from \texttt{MicrOMEGAs} ($\Omega_{\text{Micro}}h^2$) for several random points, and dedicated scans were performed to guarantee that $\DT$ works as expected for the most challenging scenarios in freeze-out, i.e.~in case of co-annihilations, resonant annihilations or when threshold effects occur. For the comparisons, we made sure to use the same total widths and input parameters as in \texttt{MicrOMEGAs}. Otherwise, differences could occur due to the different computation of the total widths used in the mediator propagators or due to the usage of different input parameters, e.g. in the (co-)annihilation channels into quark final states where we use running masses at twice the DM mass (except for the top quark mass chosen to be the pole mass). Samples of our comparisons are shown in Figs.~\ref{fig:comparison1}-\ref{fig:comparison_xsec}. \s

In Fig.~\ref{fig:comparison1}, we present the relic density $\Omega_c h^2$ as a function of the DM mass $m_\text{DM}$ for the real singlet model~\cite{McDonald:1993ex}. The Higgs sector of this model features a DM particle and a SM-like Higgs with mass 125~GeV. For the plot, we chose the mass range of the DM
particle between 5~GeV and 1000~GeV, and the freeze-out portal coupling was set to 0.1 (orange and green points) and 6 (red and black points). We observe, as expected, a resonant annihilation at $m_{\text{DM}} = m_h/2$, where $m_h$ denotes the SM-like Higgs mass value of 125~GeV, resulting in a dip in the relic density. The subsequent peaks and dips in the relic density are a consequence of the DM mass passing the thresholds for on-shell production of, respectively, a $W$ boson, $Z$ boson and finally a SM-like Higgs pair. The dip at the Higgs-mass threshold is much less pronounced for $\lambda_{\text{portal}}=0.1$ than for $\lambda_{\text{portal}} =6$, reflecting the different values of the SM-like Higgs coupling to the DM particles in the two chosen scenarios. As can be inferred from the figure, the \texttt{MicrOMEGAs} and the  $\DT$ results are in very good agreement, with differences of at most 6\%. \s

Next we compare results in the model CP in the Dark \cite{Azevedo:2018fmj,Biermann:2022meg}. It is based on an N2HDM with an imposed discrete symmetry such that the Higgs spectrum consists of a SM Higgs boson and the additional Higgs bosons, three neutral and two charged ones, residing in a dark sector with explicit CP violation. The lightest of these dark scalars must be neutral and represents the DM candidate. In Fig.~\ref{fig:comparison2}, we show the absolute value of the relative difference in the relic density between $\DT$ and \texttt{MicrOMEGAs} (in \%), defined as $|(\Omega_{\text{RelExt}} - \Omega_{\text{Micro}})|/\Omega_{\text{Micro}}$, as a function of the DM mass. The displayed points have been obtained in a scan with  \ttt{ScannerS} which checks for the most relevant theoretical and experimental
constraints, for details cf.~Refs.~\cite{Azevedo:2018fmj,Biermann:2022meg}. As can be inferred from the plot, for the bulk of the points the difference is below 2\%. \s

Larger differences can occur in models where there is an additional Higgs boson in the visible sector that can have a large total decay width, and for parameter points where we are near an $s$-channel resonance in the (co-)annihilation channel. This is shown in Fig.~\ref{fig:comparison_widths} for the dark doublet phase of the N2HDM \cite{Engeln:2020fld}. The model, which is based on a 2HDM extended by a real singlet field, features a discrete symmetry such that the Higgs spectrum consists of four dark sector particles (two neutral and two charged ones) and two visible Higgs bosons, one of which is the SM-like Higgs boson. The other non-SM-like Higgs boson, denoted by $h_{\text{non-SM}}$ in the following, can have a large total width, depending on its mass value. The DM particle is given by the lightest dark sector scalar which must be neutral. The figure shows the absolute value of the relative difference between the relic densities calculated with $\DT$ and \texttt{MicrOMEGAs}, respectively, but now with respect to $m_{h_{\text{non-SM}}}/m_{\text{DM}}$ (left) and $\Gamma_{h_{\text{non-SM}}}$ (right). 
The displayed points have been obtained from a scan with \ttt{ScannerS} and respect the most relevant theoretical and experimental constraints. We can see that the relative differences can go up to around~9\% and that the larger differences occur for parameter points where $m_{h_{\text{non-SM}}}/m_{\text{DM}}\sim2$, i.e. near an $s$-channel resonance, and for the largest $\Gamma_{h_{\text{non-SM}}}$ values. Note, however, that being near a resonance or  $\Gamma_{h_{\text{non-SM}}}$ being large does not necessarily result in a significant difference in the relic densities. The reason for the differences is due to the different implementation of the total widths in case of resonances. Our code $\DT$ makes the replacement in Eq.~(\ref{eq:propagator}) for the propagators in the amplitudes throughout the whole integration range of the c.m.~energy, with the width obtained as described in Sec.~\ref{decaywidths}. In \texttt{MicrOMEGAs}, however, the total width is taken non-zero only in a finite integration area (defined by the code) around the $s$-channel resonant point. Depending on the involved coupling strengths and total width values the differences in the results can be more or less pronounced. \s

\begin{table}[h!]
  \centering
  \begin{tabular}{lc}
    \toprule
     & N2HDM \\
    \midrule
    $m_{h_{\text{SM}}}$ & 125.09 \\
    $m_{h_{\text{non-SM}}}$ & 927.082 \\
    $m_{H_D}$ & 405.215 \\
    $m_{A_D}$ & 595.853  \\
    $m_{H_D^\pm}$ & 628.737 \\
    $\alpha$ & -0.248593 \\ 
    $v_s$ & 292.978 \\
    $m_{22}$ & 133.912 \\
    $\lambda_2$ & 3.49988 \\
    $\lambda_8$ & 2.34688 \\
    $\Gamma_{h_{\text{SM}}}$ & 0.00490257 \\
    $\Gamma_{h_{\text{non-SM}}}$ & 38.9219 \\
    \midrule
    $\Omega_{\text{RelExt}}h^2$ & 0.000314878 \\
    $\Omega_{\text{Micro}}h^2$ & 0.000339666 \\
    \bottomrule
  \end{tabular}
  \caption{Parameter point shown in Fig.~\ref{fig:comparison_xsec} for the dark doublet phase of the N2HDM. Lines 1-12 contain the input parameters, lines 13-14 the relic densities from $\DT$ and \texttt{MicrOMEGAs}, respectively. For details about the model and its input parameters, see \protect\cite{Engeln:2020fld,Azevedo:2021ylf}. For the input parameters with units, all units are in GeV.}
  \label{tab:points}
\end{table}

We exemplify this once again for the dark doublet phase of the N2HDM benchmark point defined in Tab.~\ref{tab:points}. It was chosen among the valid parameter points as a point that shows one of the largest deviations, i.e.~7\%. The SM-like Higgs boson $h_{\text{SM}}$ is the lighter of the two visible Higgs bosons, the non-SM-like one $h_{\text{non-SM}}$ with a mass of 927~GeV has a rather large total width of $\Gamma_{h_{\text{non-SM}}}\approx 39$~GeV. 
In Fig.~\ref{fig:comparison_xsec}, we compare for this benchmark point the annihilation cross section into two massive $W$ bosons, $\sigma(\text{DM DM} \to W^+ W^-)$, as a function of the c.m.~energy $\sqrt{s}$. We can see a clear deviation between $\DT$ (black points) and \texttt{MicrOMEGAs} (red points). In terms of the cross section values at specific $\sqrt{s}$ values, the maximum found deviation is $\sim100$ near the resonance $\sqrt{s}=927$~GeV. We checked that the reason for the differences is due to the different treatment of the total widths, by setting the width $\Gamma_{h_{\text{non-SM}}}$ of the additional visible sector Higgs to zero for the parameter point that we showed, resulting in a good agreement between the cross sections also outside the resonance peak.

\section{Conclusions and Next Steps \label{sec:conclusions}}
We have presented the code $\DT$, a \texttt{C++} code for the computation of the thermal relic density of SM extensions featuring a DM particle stabilized by a discrete $\mathbb{Z}_2$ symmetry. It allows for efficient parameter scans to search for parameter configurations that reproduce the relic density within the user-defined uncertainty. The code automatically computes the required amplitudes and thermally averaged cross sections based on freeze-out processes and solves the Boltzmann equation. It includes the total widths of the $s$-channel mediators to guarantee numerical stability at the thresholds. The released code includes the pre-installed models CxSM, the N2HDM in the dark doublet phase, the model CP in the Dark, the TRSM and the BDM5. The code can be applied, however, for any arbitrary model featuring a DM candidate from freeze-out. The user solely has to provide the corresponding \texttt{FeynRules} model file. We presented the usage of the code and of its search algorithms in detail in the appendix for some sample models. The code can easily be linked to other codes like e.g.~\texttt{ScannerS} to test the relevant theoretical and experimental constraints, or to \texttt{BSMPT} to trace the phase history of the model and the sensitivity of future gravitational waves experiments to possibly generated related gravitational waves. 
The code hence provides an excellent tool to efficiently test DM models w.r.t.~their capability of reproducing the measured relic density through thermal freeze-out while respecting all relevant constraints. \s

At present, the code can deal with models featuring one DM candidate. Future possible releases will aim at the inclusion of multi-component DM models or models with feebly interacting DM candidates generating the relic density through freeze-in. Further extensions shall include the computation of direct detection signals. Major upgrades finally include higher-order corrections to the freeze-out/freeze-in processes, which is currently underway for a sample model. \s

The code is publicly available and can be downloaded from the url: 
\begin{center} 
\url{https://github.com/jplotnikov99/RelExt}
\end{center}

\section*{Acknowledgements}
We thank Duarte Azevedo, Alexander Belyaev, Lisa Biermann, Pedro Gabriel, and Michael Spira for useful discussions. We thank Karo Erhardt, Kevin Schmidt, and Anasztazia Szalontai for testing the program. 
MM acknowledges financial support from the BMBF-Project 05H24VKB. KE is grateful to Avicenna-Studienwerk for financial support. JP acknowledges financial support from the Studienstiftung des Deutschen Volkes. 
RC and RS are partially supported by the Portuguese Foundation for Science and Technology (FCT) under projects no. CFTC: UIDB/00618/2020 (https://doi.org/10.54499/UIDB/00618/2020), UIDP/00618/2020, (https://doi.org/10.54499/UIDP/00618/2020)
and through the PRR (Recovery and Resilience Plan), within the scope of the investment “RE-C06-i06 - Science Plus Capacity Building”, measure “RE-C06-i06.m02 - Reinforcement of financing for International Partnerships in Science, Technology and Innovation of the PRR”, under the project with the reference 2024.03328.CERN.
RC is additionally supported by FCT with a PhD Grant No.~2020.08221.BD. We have used ChatGPT to generate a basic version of our logo which we then further modified.
\begin{appendix}
\section{Examples for Parameter Searches
\label{app:examples}}
In the following, we exemplify the usage of our code for various pre-installed models. For each of these, we first briefly introduce them and then show the sample code for the respective performed scan.

\subsection{The CxSM}
\subsubsection{The Model CxSM}

The CxSM is based on the SM with one Higgs doublet $\Phi$ extended by a complex singlet field $\mathbb{S}$ with zero isospin and hypercharge. After electroweak symmetry breaking, the doublet and singlet fields can be parametrized as
\begin{eqnarray}
\Phi = \left( \begin{array}{c}
G^+ \\ \frac{1}{\sqrt{2}} (v + H + iG^0) \end{array}
\right) \,, \quad \mathbb{S} = \frac{1}{\sqrt{2}} (v_S + S + i (v_A + A)) \;,\label{eq: cxsm_content}
\end{eqnarray}
where $H$, $S$, and $A$ are real scalar fields and $G^+$ and $G^0$ are the charged and neutral Goldstone bosons, respectively. The $v$, $v_S$, and $v_A$ are the vacuum expectation values (VEVs) of the neutral CP-even Higgs field $H$, of the CP-even singlet  field $S$ and of the CP-odd singlet field $A$, respectively. In order to incorporate a DM candidate, we require the potential to be invariant under two separate $\mathbb{Z}_2$ symmetries acting on $S$ and $A$, under which $S \to -S$ and $A\to -A$. The corresponding renormalizable potential is given by 
\begin{eqnarray}
V = \frac{m^2}{2} |\Phi|^2 + \frac{\lambda}{4} |\Phi|^4 + \frac{\delta_2}{2} |\Phi|^2 |\mathbb{S}|^2 + \frac{b_2}{2} |\mathbb{S}|^2 + \frac{d_2}{4} |\mathbb{S}|^4 + \left( \frac{b_1}{4} \mathbb{S}^2 + c.c. \right) \;,
\end{eqnarray}
where all couplings are real. Setting $v_A=0$, the $A\to -A$ symmetry is unbroken, $A$ is stable and becomes the DM candidate of the model. The other $\mathbb{Z}_2$ symmmetry is broken by $v_S \ne 0$ so that the scalars $H$ and $S$ mix. We denote the mass eigenstates obtained from the gauge eigenstates $H$ and $S$ through an orthogonal rotation by $h_i$ ($i=1,2$),
\begin{eqnarray}
\left( \begin{array}{c} h_1 \\ h_2 \end{array} \right) = R_\alpha
\left( \begin{array}{c} H \\ S 
\end{array} \right) \,.
\end{eqnarray}
The rotation matrix is given by
\begin{eqnarray}
R_\alpha=\left( \begin{array}{cc} \cos \alpha & \sin \alpha \\ - \sin\alpha & \cos \alpha \end{array} \right) \,,\label{eq: cxsm_rot}
\end{eqnarray}
and the mass eigenstates are ordered by ascending mass, i.e.~$m_{h_1} \le m_{h_2}$. The couplings of the CxSM Higgs bosons to SM particles, $g_{h_i \text{SM} \text{SM}}$, are all modified by the same coupling factor $k_i$, given by
\begin{eqnarray}
g_{h_i \text{SM} \text{SM}} = k_i
g_{H_{\text{SM}} \text{SM} \text{SM}} \;, \; k_i \equiv \left\{ \begin{array}{cc}  \cos\alpha\,, & \; i=1 \\
-\sin\alpha \, & \; i=2 \end{array} \right. \,,
\end{eqnarray}
where $g_{H_{\text{SM}} \text{SM} \text{SM}}$ denotes the SM coupling between the SM Higgs and the SM particles. We choose the following set of input parameters for our model 
\begin{eqnarray}
v\,, \; v_S\,, \; \alpha \,, \; m_{h_1} \,, \; m_{h_2} \,, \; m_A \,,
\end{eqnarray}
where $m_A$ denotes the DM mass. For the relation between these input parameters and the potential parameters,reader to \cite{Egle:2022wmq}.

\subsubsection{Code example for the CxSM}
In order to illustrate how a {\tt main.cpp} file is set up we provide a simple example for the CxSM in which we scan the DM mass and adjust the singlet VEV such that we obtain the observed relic density. Below, we show the settings block for our example:\footnote{The CxSM model file and the model files of the following examples are stored in the corresponding \texttt{FR\_modfiles} directories.}
\begin{lstlisting}[language=c++]
/* Change to desired settings starting from here
***********************************************
 */
static constexpr int MODE = 1;
static const VecString SAVEPARS = {"MA1", "MS1", "alpha", "svev"};
static const VecString CONSIDERCHANNELS = {};
VecString NEGLECTCHANNELS = {};
static const VecString NEGLECTPARTICLES = {"u", "d", "e", "mu"};
static constexpr double BEPS = 1e-6;
static constexpr double XTODAY = 1e6;
static constexpr bool FAST = true;
static constexpr bool CALCWIDTHS = true;
static constexpr bool SAVECONTRIBS = false;
/*
***********************************************
 Until here */
\end{lstlisting}
with the corresponding input file \texttt{dataInput/examples/cxsm\_example.dat}:
\begin{lstlisting}[language=bash]
MS1     |10, 1000   ,205
MA1     |10, 1000   ,50
alpha   | 0, 1.57079,0.2
svev    | 0, 1e6    ,50
\end{lstlisting}
Here, we choose mode 1 (cf.~Tab.~\ref{tab:standardsettings}) to initialize the CxSM parameters, which are the DM mass \texttt{MA1}, the second (non-SM) Higgs mass \texttt{MS1}, corresponding to $m_{h_1}$ or $m_{h_2}$, depending on its size w.r.t.~the SM-like Higgs mass of 125~GeV,  the mixing angle \texttt{alpha} between the second Higgs and the SM-like Higgs  and the VEV of the second Higgs \texttt{svev}. The mass and the VEV values of the SM-like Higgs are set in the model file. The first two values of the input file for each parameter are the boundaries between which the user can potentially scan in the respective parameters and the last value is its initial value. Since the CxSM is a Higgs portal model, we can neglect channels with light fermion final states, as the couplings are proportional to the fermion masses, which are negligible for the first two generations. Therefore, we remove the annihilation channels into up/down-quarks, electrons and muons via the \texttt{NEGLECTPARTICLES} setting, in order to speed up the computation. The setting \texttt{BEPS} is defined in Tab.~\ref{tab:standardsettings} and \texttt{XTODAY} is today's $x$ value, i.e. the $x_0$ used in the computation of the yield (see Sec.~\ref{sec:algorithms}). \s 

\begin{figure}
    \centering
    \includegraphics[width=0.6\linewidth]{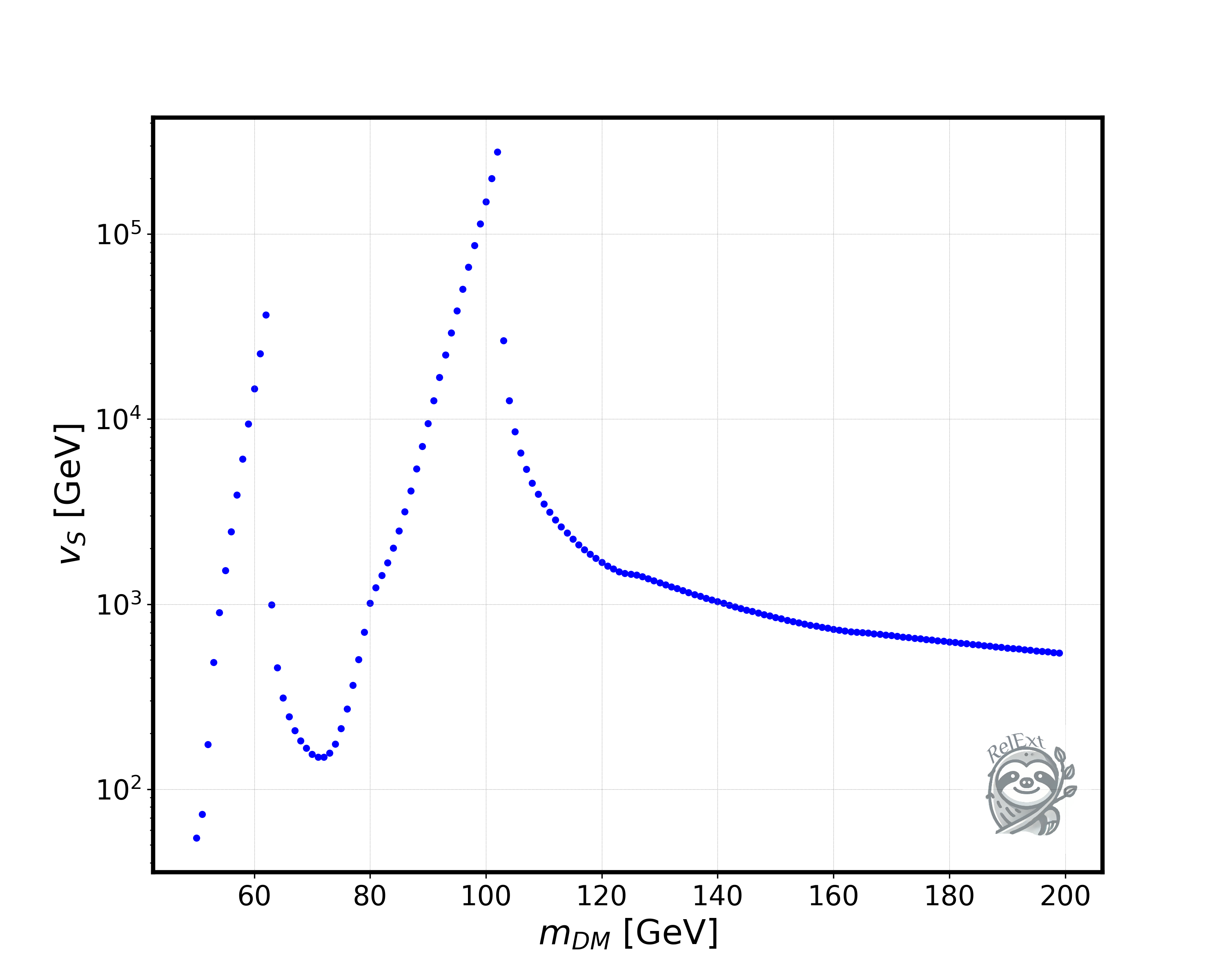}
    \caption{Singlet VEV $v_S$ over the DM mass for the CxSM. All points generate the full measured relic density within the given uncertainty.}
    \label{fig: cxsm_mass_scan}
\end{figure}

Now that we have prepared the settings we move to the main block:

\begin{lstlisting}[language=c++]
int main(int argc, char **argv) {
    clock_t begin_time = clock();
    Main M(argv, MODE, BEPS, XTODAY, FAST, CALCWIDTHS, SAVECONTRIBS);
    M.set_channels(CONSIDERCHANNELS, NEGLECTCHANNELS, NEGLECTPARTICLES);

    M.LoadParameters();
    double sav = M.GetParameter("MA1");
    for (size_t i = 0; i < 150; i++) {
        M.FindParameter("svev", 0.12, 0.001);
        M.SaveData(SAVEPARS);
        M.ChangeParameter("MA1", ++sav);
    }

    std::cout << "Computation time:\n"
              << float(clock() - begin_time) / CLOCKS_PER_SEC << "\n";
}
\end{lstlisting}
The first few lines are the same for each model. In these, the model is initialized and the (co-)annihilation channels are set. We start by loading the model parameters specified in the \ttt{cxsm\_example.dat} file via \ttt{LoadParameters}. Next, we save the DM mass \ttt{MA1=50}(GeV) into a dummy variable \ttt{sav}. In the \ttt{for} loop we first search for the value of \ttt{svev} that results in a relic density value of $\Omega h^2 = 0.120\pm0.001$ by applying the {\it second method} described in Sec.~\ref{sec:paramSearches}. Next, we call \texttt{SaveData}, which will save the previously computed relic density and the parameters specified in the setting block via \texttt{SAVEPARS}. Lastly, we add 1 GeV to the DM mass and change it via \ttt{ChangeParameter}.
During the execution of the code the output file \texttt{cxsm\_scan.dat} will be generated in the \ttt{dataOutput} folder. For our example this looks like the following:
\begin{lstlisting}[language=bash]
Omega	MA1	MS1	alpha	svev
0.11923	50	205	0.2	54.4788
0.119004	51	205	0.2	73.1971
0.119006	52	205	0.2	174.326
.
.
.
0.119907	197	205	0.2	552.402
0.119904	198	205	0.2	548.701
0.119901	199	205	0.2	545.06
\end{lstlisting}
To run the code simply call in the terminal from the \DTpath\ttt{/build} directory
\begin{lstlisting}
    ./cxsm examples/cxsm_example.dat cxsm_scan.dat
\end{lstlisting}
We can now use this data to plot \texttt{svev} over \texttt{MA1} as in Fig.~\ref{fig: cxsm_mass_scan} to illustrate the scan results. We see two peaks exactly at the thresholds of both Higgs particles, i.e.~for $m_\text{DM}=m_{h_i}/2$ ($i=1,2$). Since in this region DM annihilation becomes very efficient, the portal coupling has to decrease to ensure that the relic density is not underabundant. In the CxSM, the portal coupling is proportional to $v_S^{-1}$ and therefore $v_S$ needs to be large to compensate for the Higgs resonances.  
\subsection{The TRSM}

\subsubsection{The Model TRSM}
The previous model was a simple example containing only one dark sector particle. To illustrate how the code can be used to scan multiple parameters we consider a model with a second dark sector particle which introduces additional (co-)annihilation channels. In the TRSM we add two real singlets $\phi_2$ and $\phi_3$ to the SM which are odd under a $\mathbb{Z}_2$ symmetry. The most general renormalizable potential is given by 
\begin{align}
V_{\text{Scalar}} = &\enspace \mu_{h}^{2} |\Phi_1|^2 + \lambda_h |\Phi_1|^4 +  m_1^2 \, \phi_2^2 +  \frac{\lambda_{2}}{4!} \, \phi_2^4  + m_2^2 \, \phi_3^2 +  \frac{\lambda_{3}}{4!} \, \phi_3^4 
\nonumber
\\
+ &  \enspace  \frac{\lambda_{12}}{2} \, |\Phi_1|^2\phi_2^2  + \frac{\lambda_{13}}{2} \, |\Phi_1|^2\phi_3^2  +  \frac{\lambda_{\text{23}}}{4} \, \phi_2^2 \phi_3^2 \label{eq:pot_1_Z2}\\
+ &  \enspace m_{12}^2 \phi_2 \phi_3  + \frac{\lambda_{\text{223}}}{4} \, \phi_2^3 \phi_3 +  \frac{\lambda_{\text{233}}}{4} \, \phi_2 \phi_3^3  +  \frac{\lambda_{123}}{2} \, |\Phi_1|^2\phi_2 \phi_3   \, ,   \nonumber
\end{align}
where $\Phi_1$ is again the SM doublet as defined in Eq.~(\ref{eq: cxsm_content}) through $\Phi$. To obtain the dark sector mass eigenstates we rotate $\phi_2$ and $\phi_3$ via an orthogonal rotation matrix $R_\alpha$ in terms of the mixing angle $\alpha$,  
\begin{align}
\begin{pmatrix} \chi \\ \psi  \end{pmatrix} = 
R_\alpha \begin{pmatrix} \phi_2 \\ \phi_3  \end{pmatrix}.
\label{eq:phis}
\end{align}
Here, $\chi$ and $\psi$ are the corresponding mass eigenstates. This leaves us with the following input parameters for the model 
\begin{eqnarray}
v\,, \; \alpha \,, \; m_\chi \,, \; m_\psi, \; m_{12} \,, \; \lambda_2 \,, \; \lambda_3, \; \lambda_{12}, \; \lambda_{13}, \; \lambda_{23}, \; \lambda_{223}, \; \lambda_{233}, \; \lambda_{123}\,.
\end{eqnarray}

\subsubsection{Code example for the TRSM}
In this example, the setting block is similar to the previous one. The only change which is made, is the following
\begin{lstlisting}[language=c++]
static const VecString SAVEPARS = {"mMChi", "mMPsi", "lam12", "lam13", "lam23"};
\end{lstlisting}
The input file \texttt{dataInput/examples/trsm\_example.dat} is given by
\begin{lstlisting}
mMChi   |10, 1000   ,100
mMPsi   |10, 1000   ,120
alpha   | 0, 1.57079,0.
lam12   | 0, 40     ,1e-3
lam13   | 0, 40     ,1e-3
lam123  | 0, 40     ,1e-3
\end{lstlisting}
There are two things to note. First, we set $\alpha = 0$ such that the two dark sector particles $\phi_2$ and $\phi_3$ do not mix and correspond to the mass eigenstates $\chi$ and $\psi$, respectively. This ensures that the couplings $\lambda_{12}$, $\lambda_{13}$ and $\lambda_{123}$ can be assigned to the (co-)annihilation strength of the initial states $\chi\chi$, $\psi\psi$ and $\chi\psi$, respectively. In addition, we do not set any values for the remaining couplings, since they are purely in the dark sector and do not contribute to the relic density.\s

\begin{figure}
    \centering
    \includegraphics[width=0.7\linewidth]{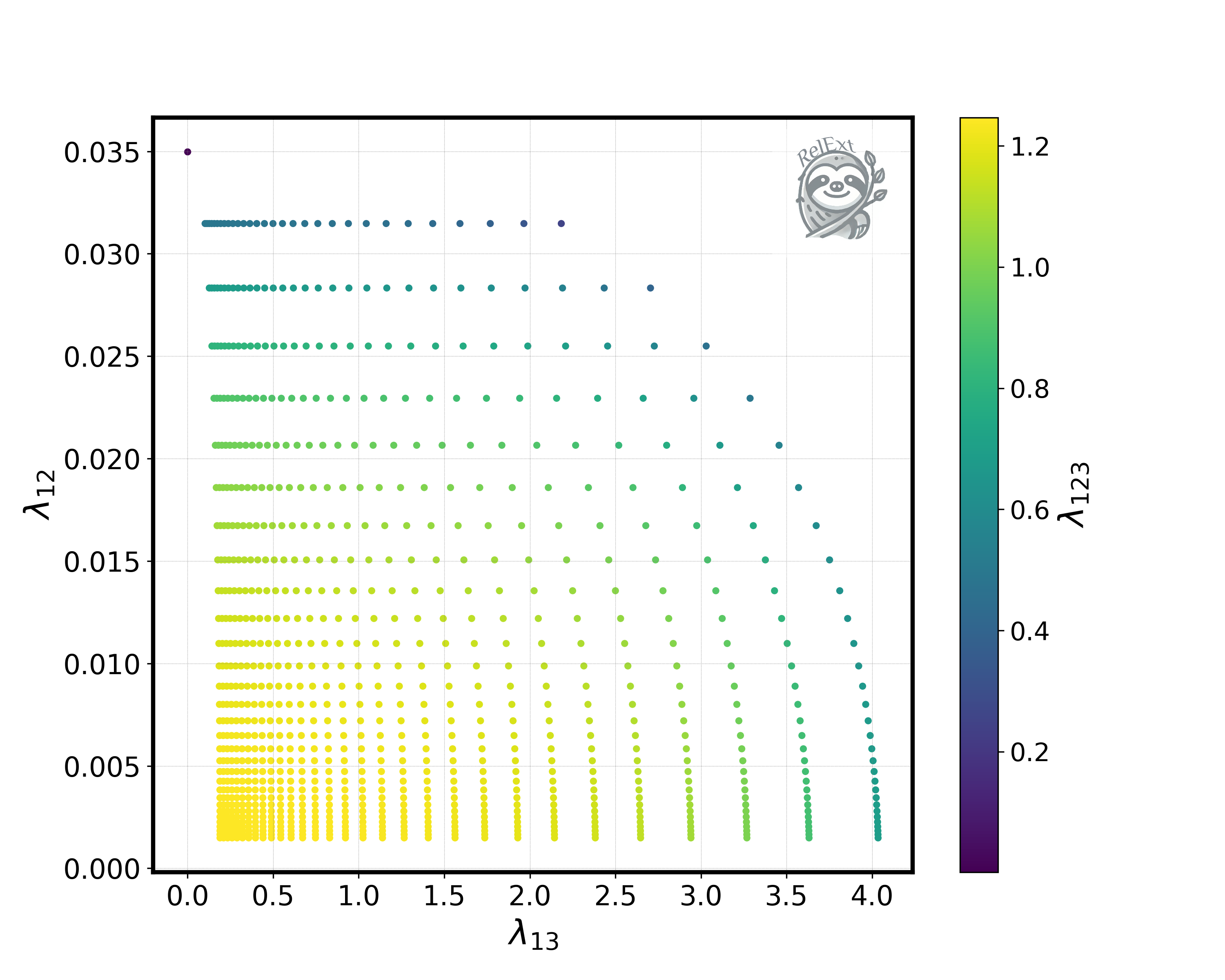}
    \caption{TRSM: Portal coupling $\lambda_{12}$, which is responsible for the annihilation of $\chi$, over the portal coupling $\lambda_{13}$ responsible for the annihilation of $\psi$. The color bar shows the portal coupling $\lambda_{123}$ responsible for the co-annihilation of $\chi$ with $\psi$. All points generate the full measured relic density within the given uncertainty.}
    \label{fig: trsm_lambda}
\end{figure}

The goal of the performed scan is to find the parameter space in which different (co-)an\-ni\-hi\-la\-tion channels dominate the relic density contribution. In order to ensure efficient (co-)an\-ni\-hi\-la\-tion, we fix the masses of the dark sector particles to $m_\chi=100$ GeV and $m_\psi=120$ GeV. The code used to perform the scan applies the {\it second method} described in Sec.~\ref{sec:paramSearches} and is given by
\begin{lstlisting}[language=c++]
int main(int argc, char **argv) {
    clock_t begin_time = clock();

    Main M(argv, MODE, BEPS, XTODAY, FAST, CALCWIDTHS, SAVECONTRIBS);
    M.set_channels(CONSIDERCHANNELS, NEGLECTCHANNELS, NEGLECTPARTICLES);

    double target = 0.12, eps = 0.001;
    M.LoadParameters();
    M.FindParameter("lam12", target, eps);
    M.SaveData(SAVEPARS);
    double save12 = M.GetParameter("lam12");
    double save13, save123 = 1e-3;
    double temp;
    for (size_t i = 0; i < 30; i++) {
        save12 *= 0.9;
        M.ChangeParameter("lam12", save12);
        M.FindParameter("lam13", target, eps);
        save13 = M.GetParameter("lam13");
        temp = save13;
        M.ChangeParameter("lam123", save123);
        for (size_t j = 0; j < 30; j++) {
            temp *= 0.9;
            M.ChangeParameter("lam13", temp);
            M.FindParameter("lam123", target, eps);
            M.SaveData(SAVEPARS);
        }
        save123 = M.GetParameter("lam123");
        M.ChangeParameter("lam123", 1e-3);
        M.ChangeParameter("lam13", save13);
    }

    std::cout << "Computation time:\n"
              << float(clock() - begin_time) / CLOCKS_PER_SEC << "\n";
}
\end{lstlisting}
We start by defining the relic density that we want to obtain, together with its allowed uncertainty by the variables \ttt{target} and \ttt{eps}. Next, we load our parameters and search for $\lambda_{12}$ such that the target relic density is obtained. In the first \ttt{for} loop that follows we reduce the value of $\lambda_{12}$ by 10\%, which leads to an increase in the relic density. To compensate for this increase we search for a $\lambda_{13}$ such that the target is obtained again and save it. We do the same procedure in the second \ttt{for} loop, but this time we reduce $\lambda_{13}$ and search for $\lambda_{123}$. After each iteration of the first loop we restore the saved parameter values. We can now plot the computed parameter space as shown in Fig. \ref{fig: trsm_lambda} with the data obtained by calling the command
\begin{lstlisting}
    ./trsm examples/trsm_example.dat trsm_scan.dat
\end{lstlisting}
from the \DTpath\ttt{/build} directory. After increasing the number of scanned points and choosing smaller variations of the couplings for each iteration, the Fig. 4 given in~\cite{Capucha:2024oaa} is reproduced.

\subsection{The CPVDM}

\subsubsection{The Model CPVDM}
The model CP in the Dark (CPVDM) \cite{Azevedo:2018fmj} is a scalar extension of the SM including two complex doublets $\Phi_1$ and $\Phi_2$ and one real singlet field $\Phi_\text{s}$. In general all the fields can acquire a non-zero vacuum expectation value after spontaneous symmetry breaking. However we will look at the scenario where only $\Phi_1$ acquires a neutral non-zero VEV $v$, such that
\begin{equation}
    \langle\Phi_1\rangle = \left(\begin{array}{c} 0\\ \frac{v}{\sqrt{2}}\end{array}\right) .
\end{equation}
After EWSB, the scalar fields can then be written as
\begin{equation}
    \Phi_1 = \left( \begin{array}{c}
G^+ \\ \frac{1}{\sqrt{2}} (v + h + iG^0) \end{array}
\right) \,,  \quad \Phi_2 = \left( \begin{array}{c}
H^+ \\ \frac{1}{\sqrt{2}} (\rho_{1} + i\eta) \end{array}
\right) \,,
\quad \Phi_\text{s} = \rho_s \;,
\end{equation}
in terms of the SM-like Higgs boson $h$, the CP-even and CP-odd fields, $\rho_i$ $(i \in \{1,s\})$ and $\eta$, respectively, the charged and neutral Goldstone bosons $G^+$ and $G^0$, and the charged field $H^+$. Furthermore, to acquire a DM candidate, a discrete $\mathbb{Z}_2$ symmetry, which takes the form 
\begin{equation}
    \Phi_1 \rightarrow \Phi_1 \, , \quad \Phi_2 \rightarrow - \Phi_2 \, , \quad \Phi_\text{s} \rightarrow -\Phi_\text{s} \; ,
\end{equation}
is applied on the Lagrangian. The most general renormalizable scalar potential respecting this symmetry, is given by
\begin{align}
  V_{\text{Scalar}} =&\enspace m_{11}^{2} \Phi_{1}^{\dagger} \Phi_{1} + m_{22}^{2} \Phi_{2}^{\dagger} \Phi_{2}
+ \dfrac{\lambda_{1}}{2} \left(\Phi_{1}^{\dagger} \Phi_{1}\right)^{2}
+ \dfrac{\lambda_{2}}{2} \left(\Phi_{2}^{\dagger} \Phi_{2}\right)^{2}\notag\\
+&\enspace \lambda_{3} \Phi_{1}^{\dagger} \Phi_{1} \Phi_{2}^{\dagger} \Phi_{2}
+ \lambda_{4} \Phi_{1}^{\dagger} \Phi_{2} \Phi_{2}^{\dagger} \Phi_{1}
+ \dfrac{\lambda_{5}}{2} \left[\left(\Phi_{1}^{\dagger} \Phi_{2}\right)^{2} + \text{h.c.}\right] \label{eq:scalpot}\\
+&\enspace \dfrac{1}{2} m_{s}^{2}\Phi_\text{s}^{2} + \dfrac{\lambda_{6}}{8} \Phi_\text{s}^{4} + \dfrac{\lambda_{7}}{2} \Phi_{1}^{\dagger} \Phi_{1} \Phi_\text{s}^{2} + \dfrac{\lambda_{8}}{2} \Phi_{2}^{\dagger} \Phi_{2} \Phi_\text{s}^{2} + (A
\Phi_{1}^{\dagger} \Phi_{2}  \Phi_{s} + \text{h.c.})\, ,\notag
\end{align}
where $m_{11}$, $m_{22}$, $m_s$ and $\lambda_i$ $(i \in [1,8])$ are real parameters, while $A$ is complex. Through the last term in Eq.~\eqref{eq:scalpot}, additional CP-violation is introduced, that is restricted solely to the dark sector. The mass eigenstates of the dark sector particles are defined as $h_i$ $(i=1,2,3)$ and are obtained via an orthogonal rotation matrix $R$, which is parameterised by the angles $\alpha_i \in [-\frac{\pi}{2}, \frac{\pi}{2}]$ $(i \in \{1,2,3\})$, 
\begin{equation}
    R=\begin{pmatrix}
    c_{\alpha_1}c_{\alpha_2}&s_{\alpha_1}c_{\alpha_2}&s_{\alpha_2}\\
        -(c_{\alpha_1}s_{\alpha_2}s_{\alpha_3}+s_{\alpha_1}c_{\alpha_3})&c_{\alpha_1}c_{\alpha_3}-s_{\alpha_1}s_{\alpha_2}s_{\alpha_3}&c_{\alpha_2}s_{\alpha_3}\\
        -c_{\alpha_1}s_{\alpha_2}c_{\alpha_3}+s_{\alpha_1}s_{\alpha_3}&-(c_{\alpha_1}s_{\alpha_3}+s_{\alpha_1}s_{\alpha_2}c_{\alpha_3})&c_{\alpha_2}c_{\alpha_3}
    \end{pmatrix}\; .\label{rotmat}
\end{equation}
Here, the notation $\sin(\alpha_i)\equiv s_{\alpha_i}$ and $\cos(\alpha_i)\equiv c_{\alpha_i}$ was employed. The mass eigenstates are then given by the relation
\begin{equation}
    \left(\begin{array}{c} h_1 \\ h_2 \\ h_3 \end{array}\right) = R  \left(\begin{array}{c} \rho_1 \\ \eta \\ \rho_s \end{array}\right) \, .
\end{equation}
Furthermore, the mass eigenstates are mass ordered, i.e., $m_{h_{1}} \leq m_{h_{2}} \leq m_{h_{3}}$. We choose the following set of parameters as input parameters for this model,
\begin{equation}
    m_{h_{1}},\quad m_{h_{2}},\quad m_{H^+},\quad \alpha_1,\quad\alpha_2,\quad\alpha_3,\quad \lambda_2,\quad \lambda_6,\quad \lambda_8,\quad m_{22},\quad m_s .
\end{equation}
For a more detailed description on the relations between the potential parameters and the chosen set of input parameters, we refer to \cite{Azevedo:2018fmj,Biermann:2022meg}.

\subsubsection{Code example for the CPVDM}\label{App: CPVDM}
The CPVDM model introduces additional parameters and co-annihilation channels with other dark sector particles compared to the simpler CxSM and TRSM. This makes the task of finding parameter regions, which generate the full relic density, more challenging. Nevertheless, we can find these regions by first observing that the model has only one Higgs portal which connects the dark sector to the SM. In such a scenario we expect a relic density dependence on the DM mass similar to the ~\ref{fig:comparison1}. We notice two distinct mass regions, one below the SM-like Higgs boson threshold and one above. With this knowledge we set up a scan for these two regions. The only changes in the settings compared to the previous two examples are
\begin{lstlisting}[language=c++]
static constexpr int MODE = 2;
static const VecString SAVEPARS = {"mH1",   "mH2",   "mHc",   "alph1", "alph2", "alph3", "m22sq", "mssq", "L2", "L6", "L8"};
\end{lstlisting}
with the parameter boundaries being set in \ttt{examples/cpvdm\_below.dat} (left) and\\ \ttt{examples/cpvdm\_above.dat} (right) as follows\\
\begin{minipage}{.5\textwidth}
\begin{lstlisting}[language=bash]
mH1     |10     ,   125
mH2     |10     ,   1000
mHc     |10     ,   1000
alph1   |0      ,   1.56
alph2   |0      ,   1.56
alph3   |0      ,   1.56
m22sq   |0      ,   1e+06
mssq    |0      ,   1e+06
\end{lstlisting}
\end{minipage}
\begin{minipage}{.5\textwidth}
\begin{lstlisting}[language=bash]
mH1     |125.1  ,   1000
mH2     |125.1  ,   1000
mHc     |125.1  ,   1000
alph1   |0      ,   1.56
alph2   |0      ,   1.56
alph3   |0      ,   1.56
m22sq   |0      ,   1e+06
mssq    |0      ,   1e+06
\end{lstlisting}
\end{minipage}
As mentioned in the previous section, the model has a mass ordering such that $m_{h_{1}} < m_{h_{2}} < m_{h_{3}}$. Furthermore, we want to make sure that the lightest dark sector particle is not charged, i.e., $m_{h1}<m_{H^\pm}$. We add these conditions in the \texttt{sources/conditions.cpp} file (which can be found in each model folder) in the following way
\begin{lstlisting}[language=C++]
bool DT::ModelInfo::check_conditions(){
    using namespace PAR;
    CHECKCONDITION(mH1 < mH2)
    CHECKCONDITION(mH2 < mH3)
    CHECKCONDITION(mH1 < mHc)

    return true;
}
\end{lstlisting}
In order to determine the relevant parameter regions, we use the {\it first method} described in Sec.~\ref{sec:paramSearches} and lay a grid on the parameter space and track the best cells of this grid. This is done with the following code,
\begin{lstlisting}[language=C++]
int main(int argc, char **argv) {
    Main M(argv, MODE, BEPS, XTODAY, FAST, CALCWIDTHS, SAVECONTRIBS);
    M.set_channels(CONSIDERCHANNELS, NEGLECTCHANNELS, NEGLECTPARTICLES);

    clock_t begin_time = clock();
    M.InitMonteCarlo(100, 500, 1, 0.12);
    for (size_t i = 1; i <= 1e5; i++) {
        M.LoadParameters();
        M.CalcRelic();
        M.SetWeight();
        M.SaveData(SAVEPARS);
    }
    std::cout << "Computation time:\n"
              << float(clock() - begin_time) / CLOCKS_PER_SEC << std::endl;
\end{lstlisting}

Here, we initialize the grid via \ttt{InitMonteCarlo} which divides each of the eight parameters specified in the input files into 100 bins of equal size.
Further, we track the 500 best cells with respect to the condition specified in the description of the first search method in Sec.~\ref{sec:paramSearches}. These will be saved in an additional output file after the scan has finished. We set the parameter \texttt{pr} of \texttt{InitMonteCarlo} to 1, forcing the code to keep generating random points rather than points from the best cells. 
\begin{figure}
    \centering\includegraphics[width=0.7\linewidth]{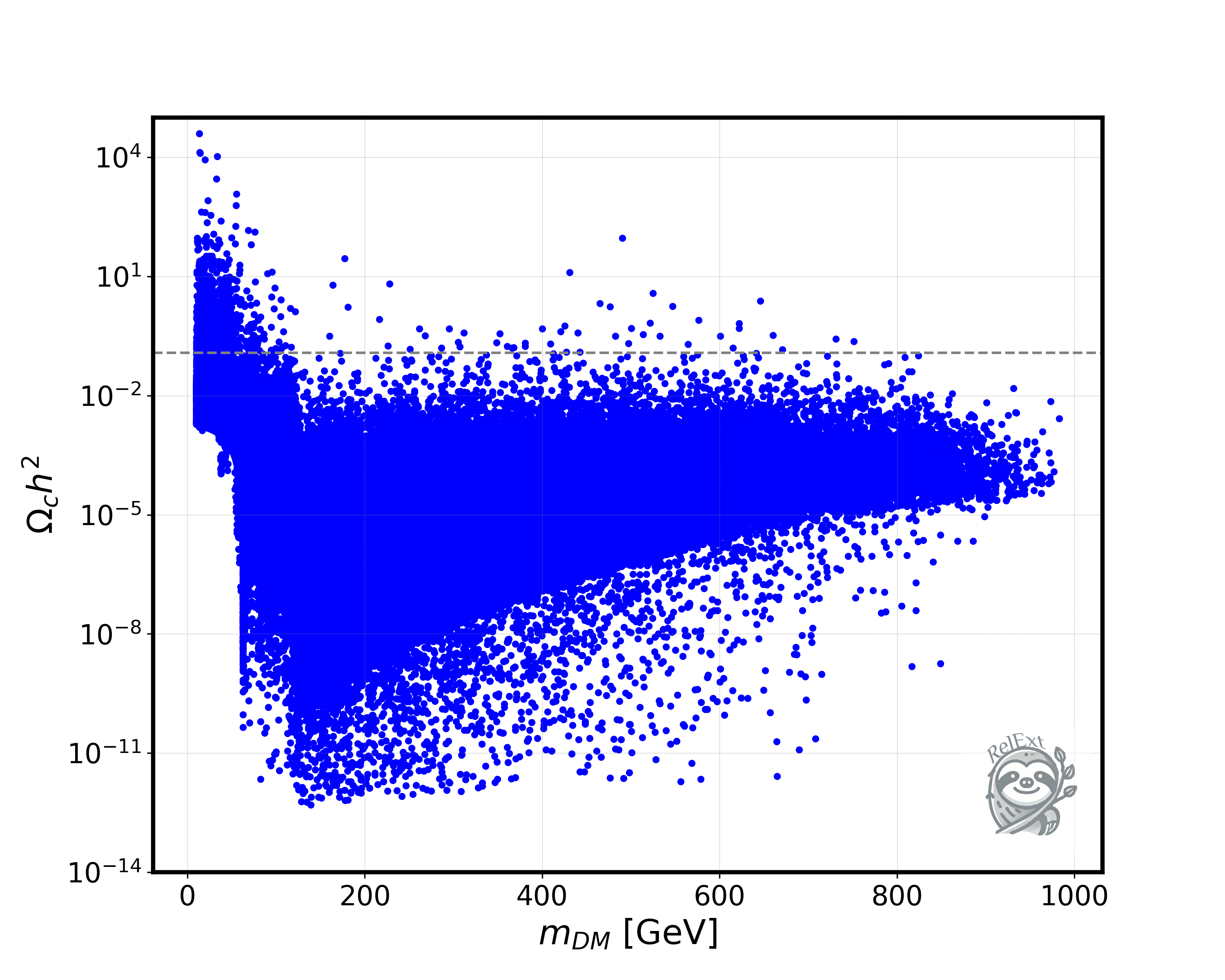}
    \caption{Computed relic density over the DM mass for the CPVDM model for $2\cdot10^5$ random parameter points. The gray dashed line shows the observed relic density of $\Omega_\text{obs}h^2=0.120$.}
    \label{fig: cpvdm_random}
\end{figure}
After running the code by calling
\begin{lstlisting}[language=bash]
    ./cpvdm examples/cpvdm_below.dat cpvdm_below.dat
    ./cpvdm examples/cpvdm_above.dat cpvdm_above.dat
\end{lstlisting}
we obtain Fig.~\ref{fig: cpvdm_random} from the generated data files, where we plot the computed relic density $\Omega_c h^2$ over the DM mass \ttt{mH1}. The gray dashed line represents the observed relic density. Out of the $2\cdot10^5$ scanned points, 27 are within the 2$\sigma$ range of the observed relic density,  $\Omega_\text{obs}h^2=0.120\pm0.002$, i.e., 0.0135\%. In this example we also tracked the computing time by using the  \texttt{clock} function.\s 

We can now generate a better sample by using the best cells which have been saved in the \texttt{dataOutput} folder, in the files \ttt{cells\_cpvdm\_below.dat} and \ttt{cells\_cpvdm\_above.dat}. For this we set \texttt{pr} in the function \ttt{InitMonteCarlo} to 0 to generate points only from the best cells and call
\begin{lstlisting}[language=bash]
    ./cpvdm ../dataOutput/cells_cpvdm_below.dat cpvdm_best_below.dat
    ./cpvdm ../dataOutput/cells_cpvdm_above.dat cpvdm_best_above.dat
\end{lstlisting}
\begin{figure}
    \centering
    \includegraphics[width=0.7\linewidth]{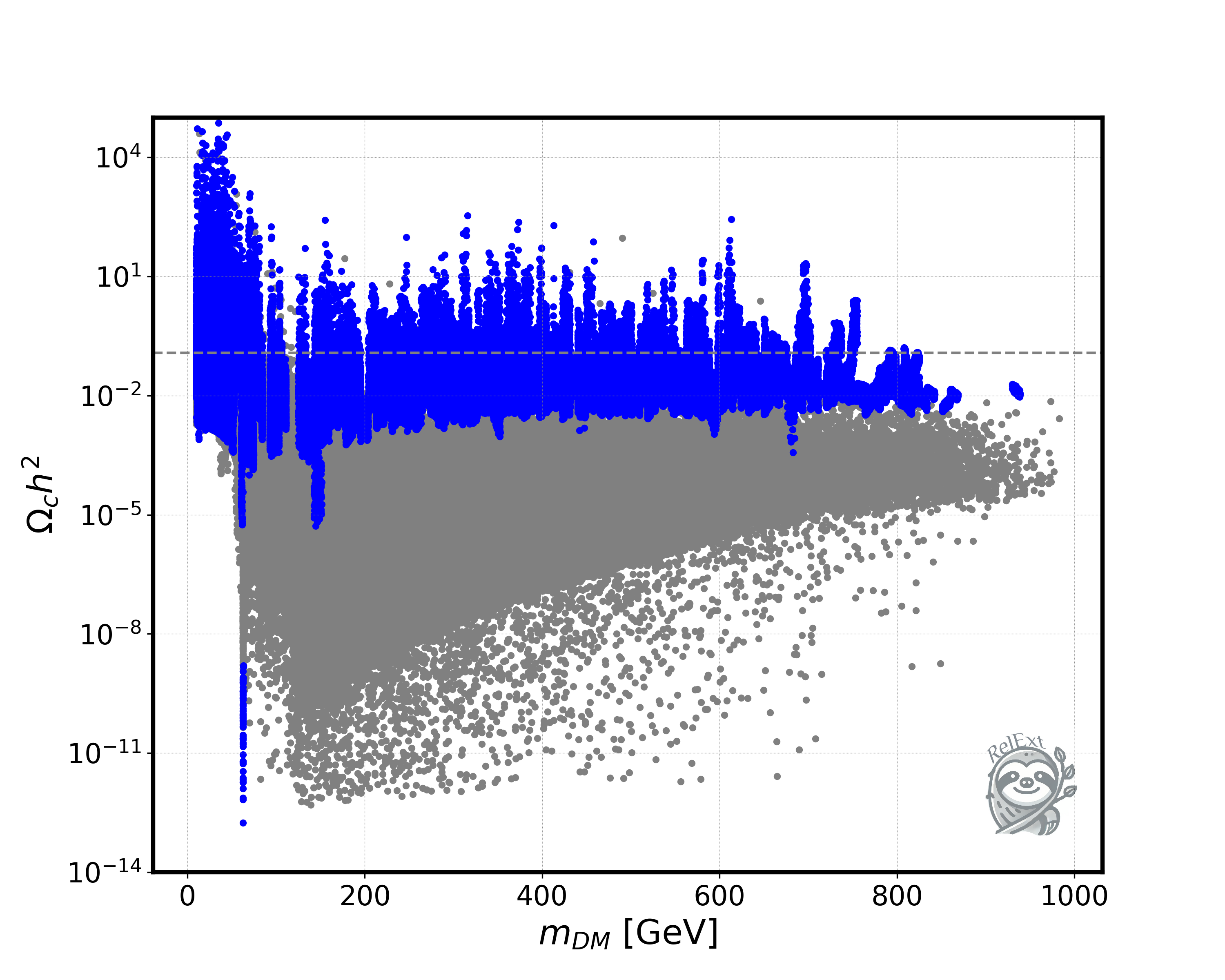}
    \caption{Computed relic density over the DM mass for the CPVDM model. The blue points are generated from the best cells of the gray points which are generated via a random scan. The gray dashed line shows the observed relic density.}
    \label{fig:cpvdm best cells}
\end{figure}
After the scan we obtain the blue points shown in Fig.~\ref{fig:cpvdm best cells}. Out of the $2\cdot10^5$ points, 3948 are within the 2$\sigma$ range of the observed relic density, i.e.~1.97\%. This means that this scan produces roughly 100 times the amount of good points compared to a simple random scan in the same amount of time as can be seen in Tab.~\ref{tab:searchcomparison}.\s 

\begin{figure}[t]
    \centering
    \includegraphics[width=0.7\linewidth]{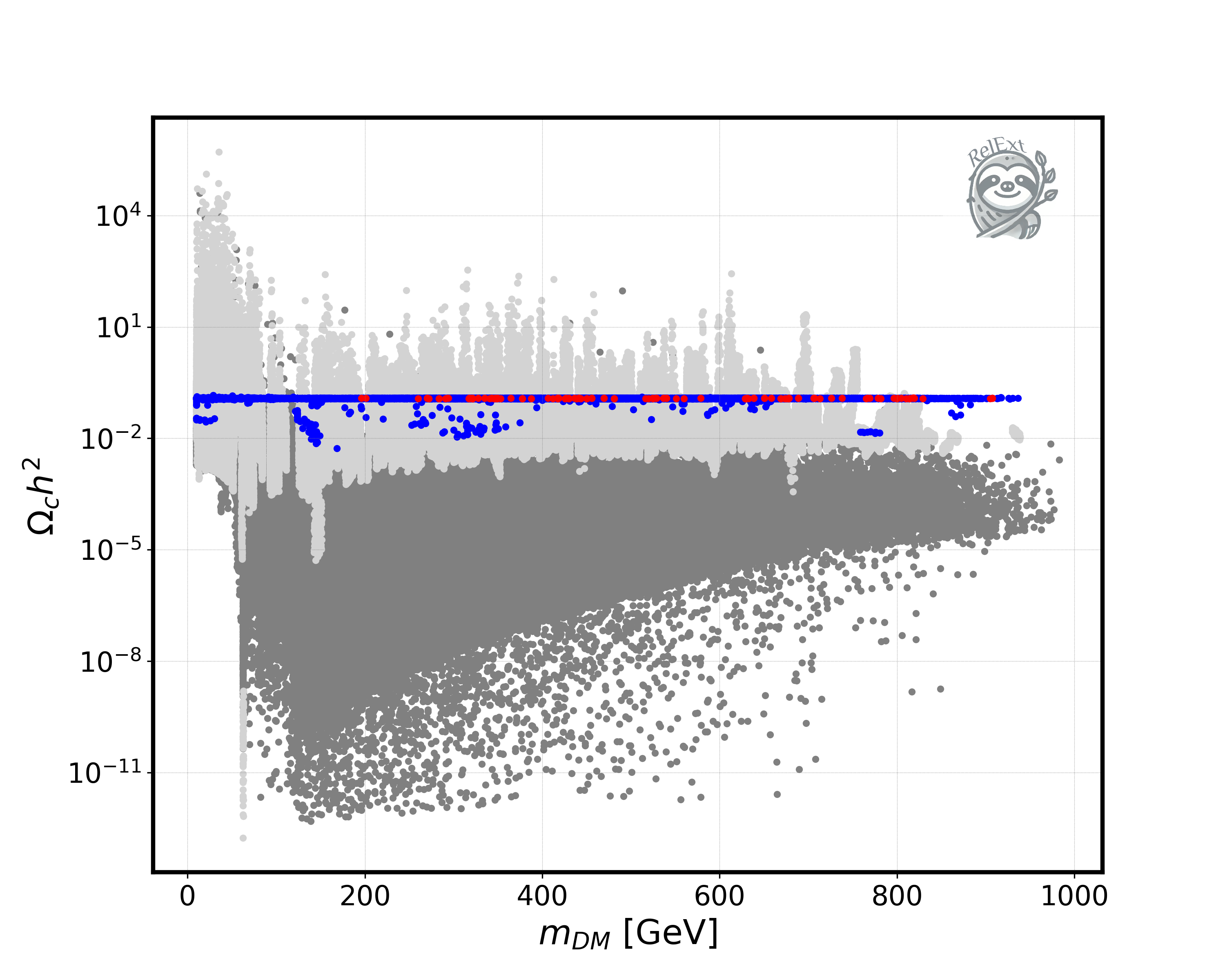}
    \caption{Computed relic density over the DM mass for the CPVDM model. The blue points are
    generated from a random scan of the best cells on which a random walk was applied. The dark gray points are random points from the entire scan region, while the light gray points are random points from the best cells. The gray dashed line shows the observed relic density. The blue points are compatible with the measured relic density within $\pm 2\sigma$, and the red points are validated against the relevant theoretical and experimental constraints using $\texttt{ScannerS}$.}
    \label{fig:cpvdm best cells rwalk}
\end{figure}
We can improve this further, by again generating random points in the best cells and then applying a random walk to obtain the observed relic density ({\it method 3} in Sec.~\ref{sec:paramSearches}). In order to do this we only need to adjust the code slightly by changing
\begin{lstlisting}
    M.CalcRelic() -> M.RWalk(0.12, 0.002, 0.01, 400)
\end{lstlisting}
and calling
\begin{lstlisting}[language=bash]
    ./cpvdm ../dataOutput/cells_cpvdm_below.dat cpvdm_bestRW_below.dat
    ./cpvdm ../dataOutput/cells_cpvdm_above.dat cpvdm_bestRW_above.dat
\end{lstlisting}

After the scan, we obtain the result shown in Fig. \ref{fig:cpvdm best cells rwalk}. We see that out of the $2\cdot10^4$ generated points, 19489 are within the 2$\sigma$ limit, i.e. 97.45\%.\footnote{The ones that are not within the 2$\sigma$ boundaries have reached the maximum step limit of the random walk.} However, due to the computational effort of the random walk algorithm the running time is much larger. For a better comparison, we need to differentiate between the two aforementioned regions. Above the Higgs threshold, the random walk produces $\sim$3.6 times more good points per second than the simple scan in the best cells. However, for the region below the Higgs threshold, the best cells scan outperforms the random walk scan by a factor $\sim$1.6. This is due to the fact that the random walk applies small changes to parameters while trying to minimize the difference between the computed relic density and the target relic density. However, in this region even small changes to the parameters can change the relic density drastically which results in a slower convergence of the algorithm. The described comparison between the methods is summarized in Tab.~\ref{tab:searchcomparison}.\s

There is, however, still a benefit in using the random walk below the Higgs threshold. The points obtained during a random walk are not bound to the best cell they originate from. This means that they can produce new best cells from the ones already found. We validated the points obtained from this procedure against the relevant theoretical and experimental constraints using \texttt{ScannerS} \cite{Muhlleitner:2020wwk,Coimbra:2013qq}. The remaining 214 points are highlighted in red in Fig. \ref{fig:cpvdm best cells rwalk}.
\begin{table}
    \centering
    \begin{tabular}{c|ccc}
        \hline
        Method & points generated & \% within $2\sigma$ & [good points]\\
        & & & /[CPU time] \\
        \hline
        \hline
        Random $m_\text{DM}>m_h$ & $10^5$ & $0.\%$ & $\sim$0 \text{s}$^{-1}$\\
        Random $m_\text{DM}<m_h$& $10^5$ & $0.027\%$ & $\sim$0.0012 \text{s}$^{-1}$\\
        \hline
        Best cells $m_\text{DM}>m_h$& $10^5$ & $0.29\%$ & $\sim$0.0178 \text{s}$^{-1}$\\
        Best cells $m_\text{DM}<m_h$& $10^5$ & $3.66\%$ & $\sim$0.135 \text{s}$^{-1}$\\
        \hline
        Best cells with \ttt{RWalk} $m_\text{DM}>m_h$& $10^4$ & $97.5\%$ & $\sim$0.065 \text{s}$^{-1}$\\
        Best cells with \ttt{RWalk} $m_\text{DM}<m_h$& $10^4$ & $97.3\%$ & $\sim$0.085 \text{s}$^{-1}$\\
        \hline
    \end{tabular}
    \caption{Total points generated by the scan (left column), fraction of points generated for which the obtained relic density is within the 2$\sigma$ limit (middle column), and number of good points (relic density within the 2$\sigma$ limit) per second (right column). The methods being compared are a random scan, a scan using only the best cells, and a scan in the best cells where a random walk is applied. The computations were performed on a liquid cooled AMD EPYC 7351 16-Core Processor machine with an x86\_64 architecture, a base clock frequency of 1.2 GHz and maximum frequency of 2.4 GHz. We used 4 cores with a total of 4 GB of RAM for our scans.}
    \label{tab:searchcomparison}
\end{table}
\subsubsection{Code example for MODE 3}
So far we have described how to generate new points within \ttt{MODE} 1 and 2, cf.~Tab.~\ref{tab:standardsettings} (the methods described in the previous sections are only compatible with these modes). However, the users can also provide their own set of parameter points for which the relic density will be computed. Here, we provide an example for the CPVDM. First, we set the mode via
\begin{lstlisting}[language=c++]
    static constexpr int MODE = 3;
\end{lstlisting}
and provide the corresponding input file, which in this example is given by
\begin{lstlisting}[language=bash]
    mH1	mH2	mHc	alph1	alph2	alph3	L2	L6	L8	mssq	m22sq
    745.97	890.30	944.876	0.0657	1.0367	0.0699	1.189	1.243	6.912	602522	656505
    .
    .
    .
    406.46	535.73	567.04	1.374	0.89	-1.0559	0.916	12.302	4.863	155598	99988
\end{lstlisting}
In the first line of the input file, the parameters of the model that we want to load in have to be listed, while the following lines contain their corresponding values. The columns in this file must be tab separated. To compute the relic density and save the result (in the file named {\tt cpvdm\_res.dat} here), we run the following command,
\begin{lstlisting}
    ./cpvdm examples/cpvdm_mode3.dat cpvdm_res.dat
\end{lstlisting}
with the \ttt{main.cpp} given below
\begin{lstlisting}[language=c++]
    int main(int argc, char **argv) {
    Main M(argv, MODE, BEPS, XTODAY, FAST, CALCWIDTHS, SAVECONTRIBS);
    M.set_channels(CONSIDERCHANNELS, NEGLECTCHANNELS, NEGLECTPARTICLES);

    clock_t begin_time = clock();
    for (size_t i = 1; i <= 1e2; i++) {
        M.LoadParameters(i);
        M.CalcRelic();
        M.SaveData(SAVEPARS);
    }
    std::cout << "Computation time:\n"
              << float(clock() - begin_time) / CLOCKS_PER_SEC << std::endl;
}
\end{lstlisting}
The key difference w.r.t.~the previous examples with the other modes is that the current row has to be provided as an input in the \ttt{LoadParameters} function. In our example this will load the parameters of the \ttt{i}'th row of the input file sample, compute the relic density and save its value. Important to note in this mode is that the \ttt{LoadParameters} function can only load the data sequentially, i.e. the index \ttt{i} can only get larger, not smaller.

\section{Possible Issues with the Generation of \ttt{FeynRules} Model Files
\label{app:issues}}
In this section, we describe some possible issues that the user may encounter when implementing a new model in \texttt{FeynRules}. If the model files for \texttt{FeynRules} are already generated, several compatibility checks should be made to ensure that $\DT$ correctly extracts the relevant information for the relic density calculation. In the following discussion, some of the terminology requires familiarity with the \texttt{FeynRules} language. A detailed description of the program can be found in~\cite{Alloul:2013bka}. The following checks/changes are the most important ones to ensure compatibility with $\DT$:

%\texttt{FeynRules} distinguishes between two types of model parameters: \ttt{external} and \ttt{internal}. External parameters correspond to free/independent parameters, while internal parameters depend on one or several of the other internal and/or external parameters of the model. Each of these parameters and their respective definitions are read by our code, and will be declared/initialized in several \texttt{C++} files. 

\begin{itemize}
    \item In $\DT$, the amplitudes relevant for freeze-out will be generated in the Feynman gauge. Some programs such as \texttt{MadGraph5\_aMC@NLO}~\cite{Alwall:2011uj} prefer to use the unitary gauge, in which case the Boolean variable \ttt{\$FeynmanGauge}, which usually appears at the top of the model files, will be set to False. In $\DT$, \ttt{\$FeynmanGauge} must be set to True to use the Feynman gauge.
    \item The value taken by the attribute \ttt{Value} of the parameter class, for external parameters should be a real number, and for internal parameters a formula written in standard \texttt{Mathematica} syntax. The user must check if the definitions of external parameters in their model (in particular the new parameters relative to the SM) follow this rule. For instance, if \ttt{alpha} is some external parameter, its \ttt{Value} cannot be another parameter, it must be a number.
    \item %the attribute \ttt{Definitions} can be used instead of \ttt{Value} for internal parameters. 
    The name and definition of each parameter is declared and initialized in $\DT$ following the same order of appearance as in the model files, thus the user must be careful to define all parameters in the correct order. For instance, all of the internal/external parameters appearing in the formula for some internal parameter must be defined prior to this parameter. 
    \item Particle classes can be used to specify the symbol and the numerical value for the masses of the different members of a particle class using the \ttt{Mass} option, as in \texttt{Mass $\to$ \{MH, Internal\}}.  In this example, \texttt{MH} is defined as an internal parameter (even though it is a free parameter), and the user needs to define \texttt{MH} again in the \texttt{M\$Parameters} list.
    \item Using several symbols to describe the same model parameter (which is sometimes done for the Yukawa couplings or the quark masses) should be avoided, as well as changing the SM part of the model files. For example, it is not uncommon to define the external variables \ttt{yms}, \ttt{ymc}, \ttt{ymb} and \ttt{ymt}, which are equal to the strange, charm, bottom and top quark masses, respectively. In this case, in order to correctly update these variables with the values for the running masses of the quarks, their names should not be changed. If they are, the user must modify the \ttt{loadparameters.cpp} file in the \ttt{\$RelExt/md\_[ModelName]/sources/} folder (see App.~\ref{app:mathematica}), or change the model files.
    \item To define and identify the particles that belong to the dark sector of the model\footnote{We stress that currently $\DT$ only supports models with a single dark sector implemented via a discrete $\mathbb{Z}_2$ symmetry.}, we use the same procedure as the code \texttt{MicrOMEGAs}, which consists of adding a tilde symbol $\sim$ before the name of the particle in the \texttt{ParticleName} and \texttt{AntiParticleName} options of particle classes. In this way, the \texttt{CalcHEP} model files automatically generated by $\DT$ can also be used in \texttt{MicrOMEGAs}.
    \item To distinguish between a scalar and a pseudoscalar particle, the user must assign specific PDG numbers to the pseudoscalar particles in the $\texttt{FeynRules}$ model files. These numbers can be 36, 46, or any three-digit number starting with 3. This can be done in the $\texttt{PDG}$ option of particle classes.
    \item The user must not use special characters in the option \ttt{ClassName} of \ttt{M\$ClassesDescription}.
    \item Some variable names are hard-coded in \ttt{FeynRules}, \ttt{FeynArts}, \ttt{FeynCalc} and \ttt{CalcHEP} and must not be used to avoid conflicts. The names of these variables are \ttt{EL}, \ttt{ee}, \ttt{gs}, \ttt{G} and \ttt{FAGS}. 
\end{itemize}  

Some of the issues mentioned will result in the abrupt termination of the code when attempting to implement a new model. If this is the case, the users will be prompted to make the necessary changes to the model files, based on the type of error message that will be printed on the terminal. The messages that they may encounter are the following:

\begin{itemize}
    \item The following SM particles are either missing from the model file or have a wrong PDG number assigned.\footnote{The standard PDG numbering scheme for the SM particles must be used, and every SM particle must be present and have a PDG number assigned.}
    \item The external parameter: \_\_ has a non numeric value given by \_\_. Please assign the parameter a numeric value or declare it as internal.
    \item Please define the \texttt{ParameterType} of \_\_ either as external or internal.
    \item The internal parameter: \_\_ has a non numeric value of \_\_ after inserting all external and previous parameters. Please order the internal parameters in the model.fr files such that they only use previously declared internal parameters or external parameters.
    \item There are no dark sector particles in your model. Please change your model.fr file and include in the \texttt{ParticleName} and \texttt{AntiParticleName} options a $\sim$ before the name of the particles that belong to this sector.
\end{itemize}

\section{\ttt{Mathematica} Code Overview
\label{app:mathematica}}
The process of a new model implementation in \ttt{RelExt} is fully automatic, but a few problems may occur, such as the ones described in App.~\ref{app:issues}, resulting in a flawed implementation. Several of these issues can be identified and fixed if the user has a deeper knowledge of the code used to set up a new model, which will be described in this section. In any case, we appreciate that any bugs/issues found are reported to us in order to patch the code. \s

The implementation of a new model involves the compilation of two files from the directory \ttt{mathematica}, namely \ttt{FR\_to\_CH\_FA.m} and \ttt{amp2cpp.m}. The first one automatically converts the \ttt{FeynRules} model files provided by the user into \ttt{FeynArts} (FA) and \ttt{CalcHEP} (CH) model files. The functions \ttt{checkPDGs[]}, \ttt{checkParameters[]} and \ttt{checkDarkSector[]} search for some of the issues mentioned in App.~\ref{app:issues}. Once all those issues are solved, the \ttt{FeynArts} and \ttt{CalcHEP} model files will be created and stored in the folders \ttt{FA\_modfiles} and \ttt{CH\_modfiles}, respectively, which can be found in the \ttt{\$RelExt/md\_[ModelName]/FR\_modfiles/} directory. Then, the amplitudes relevant for freeze-out, as well as all the necessary information to implement a new model, will be generated and stored by compiling $\texttt{amp2cpp.m}$. This file can be divided into three major blocks: amplitudes part, widths part, and mathematica to cpp part. In what follows, the most important functions, lists and segments of each block are described, starting with the amplitudes block:      

\begin{itemize}
    %\item \texttt{GSlist} list with the Goldstone fields, obtained from \texttt{FR\$GoldstoneList}. This list is used to exclude Goldstone particles from the initial/final states in the freeze-out processes.
    \item \texttt{particlelist} is the list with the FA particle identifiers (first column), their masses (second column), names (third column) and PDG numbers (fourth column). %The first three can be found in \texttt{M\$ClassesDescription}, in the \texttt{FA\_modfiles.mod} file. The PDG numbers are taken from \texttt{prtcls1.mdl}. 
    Antiparticles are included in this list, Goldstones and ghosts are excluded.
    %\item \texttt{getWidthsandDSMasses[]} function to extract the list of masses for the dark sector particles, saved in the list chmass, and all masses and widths which are not zero, saved in widths.    
    %\item \ttt{IDtoPDG[ID\_]} and \ttt{PDGtoID[PDG\_]} functions to determine the PDG number of a given FA particle identifier (\ttt{ID\_}) and vice-versa, respectively. The input of \ttt{PDGtoID[PDG\_]} is a string. For instance, if \ttt{S[1]} represents the Higgs boson, \ttt{IDtoPDG[S[1]]} gives 25, and \ttt{PDGtoID["25"]} outputs \ttt{S[1]}. 
    \item \ttt{determineDof[pID\_]} is the function to determine the degrees of freedom for a given particle, represented by its FA particle identifier (\ttt{pID\_}).
    \item \ttt{dslist}, \ttt{dsmass}, \ttt{dsnames}, \ttt{dsDof} is a set of lists that contain the dark sector FA particle identifiers, masses, names and degrees of freedom, respectively.
    \item \ttt{alldiags} are the Feynman diagrams for the \ttt{TopologyList} with all 2 $\to$ 2 processes at tree level with two dark sector particles in the initial state and none in the final state. %This list is not the final list with all the freeze-out processes.   
    \item \ttt{removeDuplicate[]} is the function to remove redundant processes from the \ttt{diagsgrouped} list. %which can be viewed by typing \ttt{diagsgrouped[[All, 1]]}. 
    A new list, \ttt{listofprocs}, is generated which, with the exception of some processes that might violate conservation laws for some internal quantum number (electric charge, lepton number, etc), is the final list with all the strictly necessary processes for freeze-out, i.e.~all the channels whose amplitudes are unique. If a wrong relic density is obtained, we recommend to check this list, in case too many or not enough channels were removed by mistake. All channels with two dark sector particles in the initial state (and none in the final state) contribute to the relic density. However, in some cases it is unnecessary to calculate their amplitude. For co-annihilations, channels with the same final state but where the particles in the initial state are swapped, also contribute to the relic density. But since these channels are exactly the same process, we only need to compute the amplitude for one of them, and the remaining one can be removed from \ttt{listofprocs}. For similar reasons, we also remove processes whose initial state is the charged conjugate of a previous one, and we include the correct multiplicative factors in the Boltzmann equation to compensate for the removed channels. %This factor is either two or four, depending on the self conjugate nature of the dark sector particles. 
    \item \ttt{breakdownAmp[proccess\_, amp\_]} is the function to identify couplings or combinations of couplings that appear multiple times in the expression of an amplitude. These quantities will be stored in tokens to save computation time (see Sec.~\ref{sec:ampl}). The input \ttt{proccess\_} is any element of the list \ttt{listofprocs}, and \ttt{amp\_} its amplitude.
    \item \ttt{calcAmps[]} is the function to determine the amplitudes for each process in \ttt{listofprocs}. The amplitudes are stored in the list \ttt{ampslist}, and the freeze-out processes in \ttt{foutlist}. %(if a process has several Feynman diagrams, the amplitudes for each are saved in different entries). 
    The auxiliary lists \ttt{coefficientlist}, \ttt{mandellist} and \ttt{tokens} will be used in the routine to tokenize the amplitudes squared. The mass of each particle participating in a given process is saved in the lists \ttt{mi}, \ttt{mj}, \ttt{mk}, \ttt{ml}. %This function is called twice. In the first call, \ttt{breakdownAmp[proccess\_, amp\_]} and subsequently \ttt{tokenize[]} will store the tokens in the lists \ttt{tokensubs} and \ttt{tokenreverse}. In the second call, \ttt{coefficientlist} will be stored as a function of tokens only, allowing us to obtain more compact expressions.
    \item \ttt{addwidthsub[den\_]}, \ttt{checkdenpol[amp\_]} are the functions used to check for $s$-channels in the amplitudes that contribute to the relic density. If any are found, the denominator of the amplitude is changed as described in Sec.~\ref{sec:ampl}. The input of \ttt{checkdenpol[amp\_]} is an element of \ttt{mandellist}, and \ttt{den\_} is the denominator of \ttt{amp\_} in string format.
    \item \ttt{relevantWs} is the list with the masses and respective total widths for the $s$-channel propagators previously found. %All masses and widths which are not zero are stored in an auxiliary list, \ttt{widths}, which is obtained from the \ttt{prtcls1.mdl} file.
    \item \ttt{calcAmp2s[]} is the function to compute the expressions of the amplitudes squared for each process in \ttt{foutlist}. These are stored in the list \ttt{final}. Depending on the number of processes, this calculation can take several minutes.
\end{itemize}

The widths part block is very similar to the previous one. The main difference is that whenever we have a scalar/pseudoscalar $s$-channel propagator, the amplitude for the total width will not be computed by \ttt{FeynCalc}, but according to the procedure described in Sec.~\ref{decaywidths}. The most important functions and lists are:

\begin{itemize}
    \item \ttt{relevantWsfields}, \ttt{relevantWidth} are the lists with the FA particle identifiers and their widths for the $s$-channel mediators relevant for freeze-out, respectively. As stated in Sec.~\ref{decaywidths}, the widths for the SM massive gauge bosons are taken from the input model files and thus must be removed from these lists. 
    \item \ttt{foutlistDecays} is the final list with all the relevant decay processes for the total widths. Each column is a FA particle identifier, with the first one representing the particle from the initial state, and the last two columns the particles from the final state.
    \item \ttt{determineType[sts\_, ind\_]} is the function to determine the type of particle appearing in the initial or final states of the decay. \ttt{sts\_} is any row of \ttt{foutlistDecays}, and \ttt{ind\_} an integer \ttt{int} which can be 1, 2 or 3 (column number of \ttt{foutlistDecays}). 
    \item \ttt{calcAmpsDecays[]} is the function to determine the amplitudes of all decays in \\\ttt{foutlistDecays}, at tree level. The types of the particles in the final states are stored in \ttt{particleType}, the amplitudes in \ttt{decayslist}, and the couplings squared for each amplitude in \ttt{couplings}.  
    \item \ttt{finalDecays}, \ttt{finalDecaysWidth} are the lists with the amplitudes squared for the decays and their partial widths, respectively. If the decaying particle is a scalar/pseudoscalar, the partial width will be set to zero by default as they are calculated in \texttt{C++} according to the procedure described in Sec.~\ref{decaywidths}. 
\end{itemize}

Finally, we have the \texttt{mathematica} to \texttt{cpp} part. In this block, the expressions that we previously obtained for the amplitudes squared and the partial widths are converted and stored in \ttt{C++} files, as well as all the necessary information about the model. The most important lists and files generated in this block are:

\begin{itemize}
    \item \ttt{inifunc[i]} is the auxiliary list which saves for each unique channel that contributes to freeze-out its \ttt{processname}, the mass and FA identifiers of each particle, and the tokenized amplitudes squared. \ttt{i} is an integer \ttt{int} which goes from 1 to the number of different initial states.
    \item \ttt{external}, \ttt{internal} are the lists with all external and internal variables of the model and their values/definitions, respectively.
    \item \ttt{neutraldsmasses} is the list with the masses of all neutral dark sector particles.
d    \item \ttt{processnameDecays}, \ttt{possibleiniDecays} are auxiliary lists that store the name of the decay processes and the name of the decaying particle, respectively.  
    \item \ttt{inifuncDecays[i]} is similar to \ttt{inifunc[i]} but for the decays. It has two extra columns for the \ttt{couplings} and \ttt{particleType} lists.
    \item \ttt{runMasses} is the list with the masses for the strange, charm, bottom and top quarks.
    \item \ttt{model.hpp}, \ttt{model.cpp}, \ttt{init.cpp}, \ttt{parametermap.cpp}, \ttt{loadparameters.cpp}, \\ \ttt{loadtokens.cpp} are the files with the declaration and initialization of all external and internal variables of the model, tokens, functions for the amplitudes squared, partial and total widths, dark sector masses, widths and masses of the relevant s-channel propagators. The first file can be found in the directory \ttt{\$RelExt/md\_[ModelName]/}, and the remaining files in \ttt{\$RelExt/md\_[ModelName]/sources/}. %In \ttt{parametermap.cpp}, a map with the names of the external variables of the model allows to recognize these variables in the data input files and to perform scans.
    If there are external variables proportional to the masses in \ttt{runMasses}, these need to be redefined in \ttt{loadparameters.cpp}, so that their values are updated when the energy scale at which the running masses are computed changes (except for the top quark, where we use the pole mass value set by the user). This happens, for instance, for the SM Yukawa couplings (see e.g. \ttt{loadparameters.cpp} in the \ttt{\$RelExt/md\_cxsm/sources/} folder).
    \item \ttt{calcwidths.cpp} is the file to calculate the running masses of the quarks at a given energy scale, and the total widths of the $s$-channel mediators. For the widths, the running masses are computed at an energy scale equal to the rest energy of the decaying particle. For DM (co-)annihilation, the energy scale is equal to twice the DM mass.
    \item \ttt{all[possibleini].cpp} are the files containing the expressions of the amplitudes squared, where \ttt{[possibleini]} are the names of the DM (co-)annihilation initial states. These are stored in \ttt{\$RelExt/md\_[ModelName]/sources/amp2s}. Channels with different initial states are saved in different files.
    \item \ttt{[possibleini]flux.cpp} are the files with the flux functions (defined in \ttt{\$RelExt/sources/\\utils.cpp}) multiplied by the amplitudes squared for each freeze-out channel. These expressions can have an additional multiplicative factor, \ttt{symfac}, to account for the redundant processes removed by the \ttt{removeDuplicate[]} is the function. The files are stored in \ttt{\$RelExt/md\_[ModelName]/sources/amp2s}.
    \item \ttt{totalW[possibleiniDecays].cpp} are the files containing the expressions of the partial widths for the relevant $s$-channel mediators, where \ttt{[possibleiniDecays]} is the name of the decaying particle. The files are stored in \ttt{\$RelExt/md\_[ModelName]/sources/amp2s}, with a different file for each mediator. For scalars/pseudoscalars, hard-coded \ttt{partial\_width} functions - which take as input the types of particles in the final states, their masses and the couplings from the amplitudes squared - are used, as described in Sec.~\ref{decaywidths}. Otherwise, we use \ttt{FeynCalc} to calculate the widths, which are set to zero if the decay is off-shell.
    \item 
    \ttt{prtcls.cpp} are the files containing the names of particles and anti-particles of the model. Particles which have no anti-particle only appear once.
\end{itemize}
\end{appendix}
%\newpage
\vspace*{0.5cm}
\bibliographystyle{h-physrev}
%\addcontentsline{toc}
\addcontentsline{toc}{section}{References}
\bibliography{main.bib}

\end{document}